\documentclass[peerreview,11pt]{IEEEtran}
\usepackage{graphicx,amsmath,amssymb,array,psfrag,setspace}
\usepackage[noadjust]{cite}
\newtheorem{theorem}{\textbf{Theorem}}

\newtheorem{definition}{\textbf{Definition}}
\newtheorem{corollary}{\textbf{Corollary}}
\usepackage{times}
\usepackage{epsf}
\usepackage{epsfig}
\usepackage{amsmath}
\usepackage{amsfonts}
\usepackage{amssymb}
\usepackage{amstext}
\usepackage{latexsym}
\usepackage{color}
\usepackage{ifthen}
\usepackage{multirow}
\usepackage{subfigure}

\newcommand{\be}{\begin{equation}}
\newcommand{\ee}{\end{equation}}
\newcommand{\bear}{\begin{eqnarray}}
\newcommand{\eear}{\end{eqnarray}}
\newcommand{\bears}{\begin{eqnarray*}}
\newcommand{\eears}{\end{eqnarray*}}
\newcommand{\bi}{\begin{itemize}}
\newcommand{\ei}{\end{itemize}}
\newcommand{\ben}{\begin{enumerate}}
\newcommand{\een}{\end{enumerate}}

\newcommand{\Expt}{\mbox{${\mathbb E}$} }

\renewcommand{\vec}[1]{\mbox{\boldmath$#1$}}

\renewcommand{\vec}[1]{\mbox{\boldmath$#1$}}

\begin{document}

\title{Side-information Scalable Source Coding}
\author{Chao Tian,~\IEEEmembership{Member,~IEEE}, Suhas N. Diggavi,~\IEEEmembership{Member,~IEEE}
}
\date{}

\maketitle

\vspace{0.5cm}
\begin{abstract}
The problem of side-information scalable (SI-scalable) source coding
is considered in this work, where the encoder constructs a progressive
description, such that the receiver with high quality side information
will be able to truncate the bitstream and reconstruct in the rate
distortion sense, while the receiver with low quality side information
will have to receive further data in order to decode.  We provide
inner and outer bounds for general discrete memoryless sources. The achievable
region is shown to be tight for the case that either of the decoders 
requires a lossless reconstruction, as well as the case with degraded
deterministic distortion measures. Furthermore we show that the gap
between the achievable region and the outer bounds can be bounded by a
constant when square error distortion measure is used. The notion of
perfectly scalable coding is introduced as both the stages operate on
the Wyner-Ziv bound, and necessary and sufficient conditions are given
for sources satisfying a mild support condition. Using SI-scalable
coding and successive refinement Wyner-Ziv coding as basic building
blocks, a complete characterization is provided for the important
quadratic Gaussian source with multiple jointly Gaussian
side-informations, where the side information quality does not have to
be monotonic along the scalable coding order. Partial result is
provided for the doubly symmetric binary source with Hamming
distortion when the worse side information is a constant, for which
one of the outer bound is strictly tighter than the other one.
\end{abstract}

\section{Introduction}
\label{sec:intro}

Consider the following scenario where a server is to broadcast
multimedia data to multiple users with different side informations,
however the side informations are not available at the server. A user
may have such strong side information that only minimal additional
information is required from the server to satisfy a fidelity
criterion, or a user may have barely any side information and expect
the server to provide virtually everything to satisfy a (possibly
different) fidelity criterion.

A naive strategy is to form a single description and broadcast it to
all the users, who can decode only after receiving it completely
regardless of the quality of their individual side
informations. However, for the users with good-quality side
information (who will simply be referred to as the good users), most
of the information received is redundant, which introduces a delay
caused simply by the existence of users with poor-quality side
informations (referred to as the bad users) in the network. It is
natural to ask whether an opportunistic method exists, {\em i.e.,} whether
it is possible to construct a two-layer description, such that the
good users can decode with only the first layer, and the bad users
receive both the first and the second layer to reconstruct. Moreover,
it is of importance to investigate whether such a coding order
introduces any performance loss. We call this coding strategy {\em
side-information scalable} (SI-scalable) source coding, since the
scalable coding direction is from the good users to the bad users. In
this work, we consider mostly two-layer systems, except the quadratic
Gaussian source for which the solution to the general multi-layer
problem is given.

This work is related to the successive refinement problem, where a
source is to be encoded in a scalable manner to satisfy different
distortion requirement at each individual stage. This problem was
studied by Koshelev \cite{Koshelev:80}, and by Equitz and Cover
\cite{EquitzCover:91}; a complete characterization of the rate-distortion region can be found in  \cite{Rimoldi:94}. Another
related problem is the rate-distortion for source coding with side
information at the decoder \cite{WynerZiv:76}, for which Wyner and Ziv
provided conclusive result (now widely known as the Wyner-Ziv
problem).  Steinberg and Merhav \cite{SteinbergMerhav:04} recently
extended the successive refinement problem in the Wyner-Ziv setting
(SR-WZ), when the second
stage side information $Y_2$ is better than that of the first stage
$Y_1$, in the sense that $X\leftrightarrow Y_2\leftrightarrow Y_1$
forms a Markov string. The extension to multistage systems with
degraded side informations in such a direction was recently completed
in \cite{TianDiggavi:05}.  Also relevant is the work by Heegard and
Berger \cite{HeegardBerger:85} (see also \cite{Kaspi:94}), where the
problem of source coding when side information may be present at the
decoder was considered; the result was extended to the multistage case
when the side informations are degraded. This is quite similar to the
problem being considered here and in \cite{SteinbergMerhav:04}\cite{TianDiggavi:05}, however without the scalable coding requirement.

\begin{figure}[tb]
  \centering 
\includegraphics[scale=0.6]{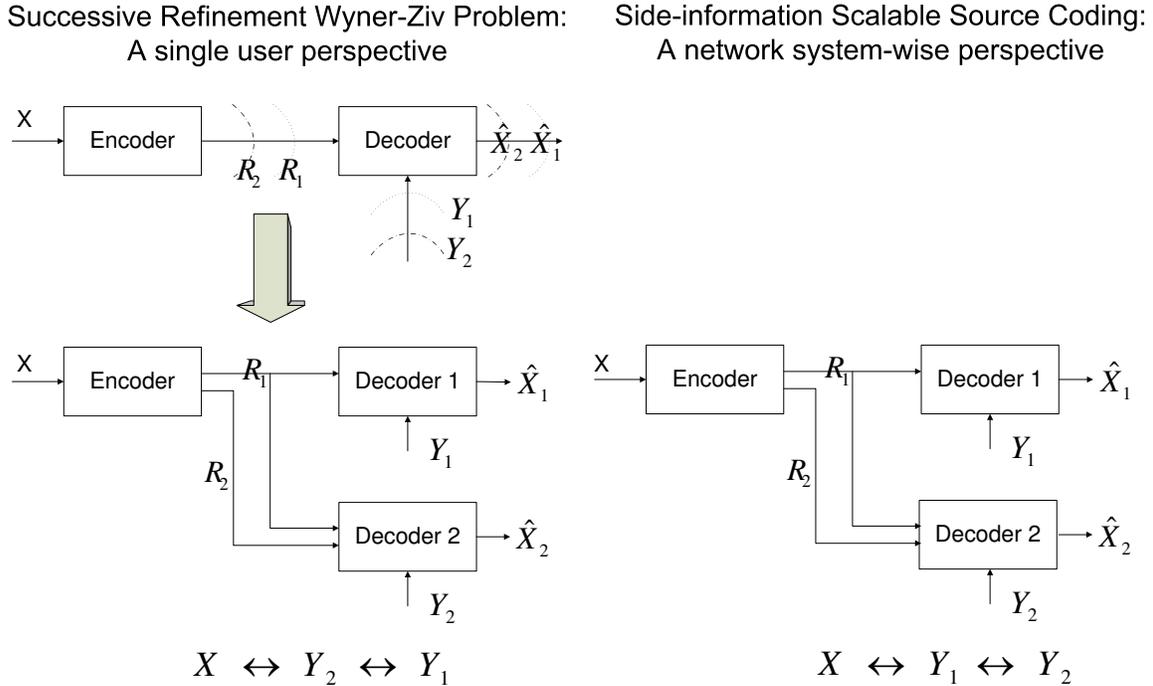}
\caption{The SR-WZ system vs. the SI-scalable system. \label{fig:SRWZ}}
\end{figure}

Both the SR-WZ \cite{SteinbergMerhav:04}\cite{TianDiggavi:05} and
SI-scalable problems can be thought as special cases of the problem of
scalable source coding with no specific structure imposed on the
decoder SI; this general problem appears to be quite difficult, since even without the scalable requirement, a complete solution to the problem has not been found \cite{HeegardBerger:85}. Here we emphasize that the SR-WZ and the SI-scalable problem are quite different in terms of their
applications, though they seem similar since only the order of SI
quality that is reversed.  Roughly speaking, in the SI-scalable
problem, the side information $Y_2$ at the later stage is worse than
the side information $Y_1$ at the early stage, while in the SR-WZ
problem, the order is reversed. In more mathematically precise terms,
for the SI-scalable problem, the side informations are degraded as
$X\leftrightarrow Y_1\leftrightarrow Y_2$, in contrast to the SR-WZ
problem where the reversed order is specified as $X\leftrightarrow
Y_2\leftrightarrow Y_1$. The two problems are also different in terms
of their possible applications. The SR-WZ problem is more applicable
for a single server-user pair, when the user is receiving side
information through another channel, and at the same time receiving
the description(s) from the server; for this scenario, two decoders
can be extracted to provide a simplified model. On the other hand, the
SI-scalable problem is more applicable when multiple users exist in
the network, and the server wants to provide a scalable description,
such that the good user is not jeopardized unnecessarily (see
Fig. \ref{fig:SRWZ}).

It is also worth pointing out that Heegard and Berger showed when the
scalable coding requirement is removed, the optimal encoding by itself
is in fact naturally progressive from the bad user to the good one; as
such, the SI-scalable problem is expected to be more difficult than
the SR-WZ problem, since the encoding order is reversed from the
natural one.  This difficulty is encapsulated by the fact that in the
SR-WZ ordering the decoder with better SI is able to decode whatever
message was meant for the decoder with worse SI and hence the first
stage can be maximally useful. However, in the SI-scalable problem an
additional tension exists in the sense that the second-stage decoder
will need extra information to disambiguate the information of the
first stage.


The problem is well understood for the lossless case. The key
difference from the lossy case is that the quality of the side
informations can be naturally determined by the value of $H(X|Y)$. By
the seminal work of Slepian and Wolf \cite{slepianwolf:73}, $H(X|Y)$
is the minimum rate of encoding $X$ losslessly with side information
$Y$ at the decoder, thus in a sense a larger $H(X|Y)$ corresponds to
weaker side information. If $H(X|Y_1)<H(X|Y_2)$, then the rate
$(R_1,R_2)=(H(X|Y_1),H(X|Y_2)-H(X|Y_1))$ is achievable, as noticed by
Feder and Shulman \cite{FederShulman:02}. Extending this observation
and a coding scheme in \cite{Csiszarkorner}, Draper \cite{Draper:04}
proposed a universal incremental Slepian-Wolf coding scheme when the
distribution is unknown, which inspired Eckford and Yu
\cite{EckfordYu:05} to design rateless Slepian-Wolf LDPC code.  For
the lossless case, there is no loss of optimality by using a scalable
coding approach; an immediate question is to ask whether the same is
true for the lossy case in terms of rate distortion, which we will
show to be not so in general. In this rate-distortion setting, the
order of goodness by the value of $H(X|Y)$ is not sufficient because
of the presence of the distortion constraints. This motivates the
Markov condition $X\leftrightarrow Y_1\leftrightarrow Y_2$ introduced
for the SI-scalable coding problem.  Going further along this point of
view, the SI-scalable problem is also applicable in the single user
setting, when the source encoder does not know exactly which side
information the receiver has within a given set. Therefore it can be
viewed as a special case of the side-information universal rate
distortion coding.

In this work, we formulate the problem of side information scalable
source coding, and provide two inner bounds and two outer bounds for
the rate-distortion region. One of the inner-bounds has the same
distortion and rate expressions as one of the outer bounds, and they
differ in the domain of optimization only by a Markov string
requirement. Though the inner and the outer bounds do not coincide in
general, the inner bounds are indeed tight for the case when either
the first stage or the second stage requires a lossless
reconstruction, as well as for the case when certain deterministic distortion measures are taken. Furthermore, a conclusive result is given for the
quadratic Gaussian source with any finite number of stages and arbitrary
correlated Gaussian side informations.

With this set of inner and outer bounds, the problem of {\em perfect
scalability} is investigated, defined as when both of the layers can
achieve the corresponding Wyner-Ziv bounds; this is similar to the
notion of (strict) successive refinability in the SR-WZ problem
\cite{SteinbergMerhav:04}\cite{TianDiggavi:05}\footnote{In the rest of
the paper, decoder one, respectively decoder two, will also be
referred to as the first stage decoder, respectively second stage
decoder, depending on the context.}. Necessary and sufficient
conditions are derived for general discrete memoryless sources to be
perfectly scalable under a mild support condition. By using the tool
of rate-loss introduced by Zamir \cite{Zamir:96A}, we further show that
the gap between the inner bounds and the outer bounds are bounded by a
constant when squared error distortion measure is used, and thus the
inner bounds are ``nearly sufficient", in the sense as given in
\cite{Lastras:06}.


In addition to the result for the Gaussian source, partial result is
provided for the doubly symmetric binary source (DSBS) with Hamming
distortion measure when the second stage does not have side
information, for which the inner bounds and outer bounds coincide in
certain distortion regimes. It is shown one of the outer bound can be
strictly better than the other for this source.

The rest of the paper is organized as follows.  In Section \ref{sec:prelim} we
define the problem and establish the notation.  In Section
\ref{sec:Ach}, we provide inner and outer bounds to the
rate-distortion region and show that the bounds coincide in certain special cases.  The notion of
perfectly scalable is introduced in Section \ref{sec:perfScalable}
together with the example of a binary source. The rate loss method is
applied in Section \ref{sec:rateloss} to show the gap between the
inner bound and the outer bounds is bounded. In \ref{sec:Gaussian},
the Gaussian source is treated within a more general setting. We
conclude the paper in Section \ref{sec:disc}.

\section{Notation and Preliminaries}
\label{sec:prelim}



Let $\mathcal{X}$ be a finite set and let $\mathcal{X}^n$ be the set
of all $n$-vectors with components in $\mathcal{X}$.  Denote an
arbitrary member of $\mathcal{X}^n$ as $x^n=(x_1,x_2,\ldots  ,x_n)$, or
alternatively as $\vec{x}$. Upper case is used for random variables and vectors. A
discrete memoryless source (DMS) $(\mathcal{X},P_X)$ is an infinite
sequence $\{X_i\}_{i=1}^{\infty}$ of independent copies of a random variable
$X$ in $\mathcal{X}$ with a generic distribution $P_X$ with
$P_X(x^n)=\prod_{i=1}^nP_X(x_i)$.  
Similarly, let $(\mathcal{X},\mathcal{Y}_1,\mathcal{Y}_2, P_{XY_1Y_2})$ be a discrete memoryless
three-source with generic distribution $P_{XY_1Y_2}$; the subscript will be dropped when it is clear from the context as $P(X,Y_1,Y_2)$.

Let $\hat{\mathcal{X}}_1$ and $\hat{\mathcal{X}}_2$ be finite reconstruction alphabets.
Let $\mathit{d}_j:\mathcal{X}\times\hat{\mathcal{X}}_j\rightarrow
\left[0,\infty\right)$, $j=1,2$ be two distortion measures. 
The single-letter distortion extension of $d_j$ to vectors is defined as
\begin{eqnarray}
\mathit{d}_j(\vec{x},\vec{\hat{x}})=\frac{1}{n}\sum_{i=1}^n
\mathit{d}_j(x_i,\hat{x}_{i}), \quad \forall \vec{x}\in \mathcal{X}^n,
\quad \vec{\hat{x}}\in \hat{\mathcal{X}}_j^n, \quad\quad j= 1,2.
\end{eqnarray}

\begin{definition}
An $(n,M_1,M_2,D_1,D_2)$ rate distortion (RD) SI-scalable code for
source $X$ with side information $(Y_1,Y_2)$ consists of two encoding
functions $\phi_i$ and two decoding functions $\psi_i$,
$i=1,2$:
\begin{eqnarray}
\phi_1:\mathcal{X}^n\rightarrow I_{M_1},&\quad\quad&
\phi_2:\mathcal{X}^n\rightarrow I_{M_2},\\
\psi_1:I_{M_1}\times \mathcal{Y}_1^n \rightarrow \hat{\mathcal{X}}_1^n,&\quad\quad&
\psi_2:I_{M_1}\times I_{M_2} \times\mathcal{Y}_2^n \rightarrow \hat{\mathcal{X}}_2^n,
\end{eqnarray}
where $I_k=\{1,2,\ldots  ,k\}$, such that 
\begin{eqnarray}
\Expt \mathit{d}_1(X^n,\psi_1(\phi_1(X^n),Y_1^n))\leq D_1,\\
\Expt \mathit{d}_2(X^n,\psi_2(\phi_1(X^n),\phi_2(X^n),Y_2^n))\leq D_2,
\end{eqnarray}
where $\Expt$ is the expectation operation. 
\end{definition}

\begin{definition}
A rate pair $(R_1,R_2)$ is said to be $(D_1,D_2)$-achievable for
SI-scalable encoding with side information $(Y_1,Y_2)$, if for any
$\epsilon>0$ and sufficiently large $n$, there exist an
$(n,M_1,M_2,D_1+\epsilon,D_2+\epsilon)$ RD SI-scalable code, such that
$R_1+\epsilon \geq \frac{1}{n}\log(M_1)$ and $R_2+\epsilon \geq
\frac{1}{n}\log(M_2)$.
\end{definition}

Denote the collection of all the $(D_1,D_2)$-achievable rate pair
$(R_1,R_2)$ for SI-scalable encoding as $\mathcal{R}(D_1,D_2)$, and we
seek to characterize this region when $X\leftrightarrow
Y_1\leftrightarrow Y_2$ forms a Markov string (see similar but
different degradedness conditions in \cite{SteinbergMerhav:04, TianDiggavi:05}). The Markov condition in effect specifies the
{\em goodness} of the side informations.

The rate-distortion function for degraded side-informations was
established in \cite{HeegardBerger:85} for the non-scalable coding
problem. In light of the discussion in Section \ref{sec:intro}, it
gives a lower bound on the sum-rate for any RD SI-scalable code. More
precisely, in order to achieve distortion $D_1$ with side information
$Y_1$, and achieve distortion $D_2$ with side information $Y_2$, when
$X\leftrightarrow Y_1\leftrightarrow Y_2$, the rate-distortion
function is
\begin{eqnarray}
R_{HB}(D_1,D_2)=\min_{p(D_1,D_2)}[I(X;W_2|Y_2)+I(X;W_1|W_2,Y_1)] ,
\end{eqnarray}
where $p(D_1,D_2)$ is the set of all random variable $(W_1,W_2)\in
\mathcal{W}_1\times\mathcal{W}_2$ jointly distributed with the generic
random variables $(X,Y_1,Y_2)$, such that the following conditions are
satisfied\footnote{This form is slightly different from the one in
\cite{HeegardBerger:85} where $f_1$ was defined as $f_1(W_1,W_2,Y)$,
but it is straightforwardly to verify that they are equivalent. The
cardinality bound is also ignored, which is not essential here.}:
({\sf i}) $(W_1,W_2)\leftrightarrow X \leftrightarrow
Y_1\leftrightarrow Y_2$ is a Markov string; ({\sf ii})
$\hat{X}_1=f_1(W_1,Y_1)$ and $\hat{X}_2=f_2(W_2,Y_2)$ satisfy the
distortion constraints. Notice that the rate distortion function
$R(D_1,D_2)$ given above suggests an encoding and decoding order from
the bad user to the good user.

Wyner and Ziv \cite{WynerZiv:76} showed that under the following quite
general assumption that the distortion measure is chosen in the set
$\Gamma_d$ defined as
\begin{eqnarray}
\label{eq:GammaDef}
\Gamma_d\stackrel{\Delta}{=}\{\mathit{d}(\cdot,\cdot):
\mathit{d}(x,x)=0,\mbox{and } \mathit{d}(x, \hat{x})>0 \mbox{ if }
\hat{x}\neq x\},
\end{eqnarray}
then the rate distortion function satisfies $R_{X|Y}^*(0)=H(X|Y)$,
where $R_{X|Y}^*(D)$ is the well-known Wyner-Ziv rate distortion
function with side information $Y$.  If the same assumption is made on
the distortion measure $d_1(\cdot,\cdot)\in\Gamma_d$, then we can
easily show (using an argument similar to the remark (3) in
\cite{WynerZiv:76}) that
\begin{eqnarray}
R_{HB}(0,D_2)=\min_{p(D_2)}[I(X;W_2|Y_2)+H(X|W_2,Y_1)], 
\label{eqn:HBcoro}
\end{eqnarray}
where $p(D_2)$ is the set of all random variable $W_2$ such that
$W_2\leftrightarrow X \leftrightarrow Y_1\leftrightarrow Y_2$ is a
Markov string, and $\hat{X}_2=f_2(W_2,Y_2)$ satisfies the distortion
constraint.

\section{Inner and Outer Bounds}
\label{sec:Ach}

To provide intuition into the the SI-scalable problem, we first
examine a simple Gaussian source under the mean squared error (MSE)
distortion measure, and describe the coding schemes informally.

Let $X\sim \mathcal{N}(0,\sigma_x^2)$ and $Y_1=Y=X+N$, where $N\sim
\mathcal{N}(0,\sigma_N^2)$ is independent of $X$; $Y_2$ is simply a
constant, {\em i.e.,} no side information at the second decoder.
$X\leftrightarrow Y_1\leftrightarrow Y_2$ is indeed a Markov
string. To avoid lengthy discussion on degenerate regimes, assume
$\sigma_N^2\approx\sigma_x^2$, and consider only the following extreme
cases.
\begin{itemize}
\item $\sigma_x^2\gg D_1 \gg D_2$: It is known binning with a Gaussian
codebook, generated using a single-letter mechanism ({\em i.e.,} as an 
i.i.d. product distribution of the single-letter form) as $W_1=X+Z_1$,
where $Z_1$ is a zero-mean Gaussian random variable independent of $X$
such that $D_1=\Expt[X-\Expt(X|Y,W_1)]^2$, is optimal for Wyner-Ziv
coding. This coding scheme can still be used for the first stage. In
the second stage, by direct enumeration in the list of possible
codewords in the particular bin specified in the first stage, the
exact codeword can be recovered by decoder two, who does not have any
side information. Since $\sigma_x^2\gg D_1 \gg D_2$, $W_1$ alone is
not sufficient to guarantee a distortion $D_2$, {\em i.e.,} $D_2 \ll
\Expt[X-\Expt(X|W_1)]^2$. Thus a successive refinement codebook, say
using a Gaussian random variable $W_2$ conditioned on $W_1$ such that
$D_2=\Expt[X-\Expt(X|W_1,W_2)]^2$, is needed. This leads to the
achievable rates:
\begin{eqnarray}
R_1\geq I(X;W_1|Y),\quad R_1+R_2\geq
I(X;W_1|Y)+I(W_1;Y)+I(X;W_2|W_1)=I(X;W_1,W_2).
\end{eqnarray}

\item $\sigma_x^2 \gg D_2 \gg D_1 $: If we choose $W_1=X+Z_1$ such
that $D_1=\Expt[X-\Expt(X|Y,W_1)]^2$ and use the coding method in the
previous case, then since $D_2 \gg D_1$, $W_1$ is sufficient to
achieve distortion $D_2$, {\em i.e.,}
$D_2\gg\Expt[X-\Expt(X|W_1)]^2$. The rate needed for the enumeration
is $I(W_1;Y)$, and it is rather wasteful since $W_1$ is more than we
need. To solve this problem, we construct a coarser description using
random variable $W_2=X+Z_1+Z_2$, such that
$D_2=\Expt[X-\Expt(X|W_2)]^2$. The encoding process has three
effective layers for the needed two stages: ({\sf i}) the first layer
uses Wyner-Ziv coding with codewords generated by $P_{W_2}$ ({\sf ii})
the second layer uses successive refinement Wyner-Ziv coding with
$P_{W_1|W_2}$ ({\sf iii}) the third layer enumerates the specific
$W_2$ codeword within the first layer bin. Note that the first two
layers form a SR-WZ scheme with identical side information $Y$ at the
decoder. For decoding, decoder one decodes the first two layers with
side information $Y$, while decoder two decodes the first and the
third layer without side information. By the Markov string
$X\leftrightarrow W_1 \leftrightarrow W_2$, this scheme gives the
following rates:
\begin{eqnarray}
R_1&\geq& I(X;W_1,W_2|Y)=I(X;W_1|Y)\nonumber\\
R_1+R_2&\geq& I(X;W_1|Y)+I(W_2;Y)=I(X;W_2)+I(X;W_1|Y,W_2).
\end{eqnarray} 
\end{itemize} 

It is seen in the above discussion the specific coding schemes depend
on the distortion values, which is not desirable since this usually
suggests difficulty in proving the converse. The two coding schemes
can be unified into a single one by introducing an auxiliary random
variable, as will be shown in the sequel, however, it appears the
converse is indeed quite difficult to prove.
 
In the rest of this section, inner and outer bounds for
$\mathcal{R}(D_1,D_2)$ are provided. The coding schemes for the above
Gaussian example are naturally generalized to give the inner
bounds. It is further shown that the inner bounds are in fact tight for certain special cases.

\subsection{Two inner bounds}

Define the region $\mathcal{R}_{in}(D_1,D_2)$ to be the set of all
rate pairs $(R_1,R_2)$ for which there exist random variables
$(W_1,W_2,V)$ in finite alphabets
$\mathcal{W}_1,\mathcal{W}_2,\mathcal{V}$ such that the following
condition are satisfied.

\begin{enumerate}\label{enm:conditions1}
\item $(W_1,W_2,V)\leftrightarrow X \leftrightarrow Y_1
\leftrightarrow Y_2$ is a Markov string.
\item There exist deterministic maps 
$f_j: \mathcal{W}_j\times \mathcal{Y}_j \rightarrow \hat{\mathcal{X}_j}$
such that
\begin{eqnarray}
\Expt\mathit{d}_j(X,f_j(W_j,Y_j))\leq D_j, \quad j=1,2.
\end{eqnarray}
\item The non-negative rate pairs satisfy:
\begin{eqnarray}
\label{eqn:rates1}
R_1\geq I(X;V,W_1|Y_1), \quad
R_1+R_2\geq I(X;V,W_2|Y_2)+I(X;W_1|Y_1,V).
\end{eqnarray} 
\item $W_1\leftrightarrow (X,V) \leftrightarrow W_2$ is a Markov string.
\item The alphabets $\mathcal{V}$, $\mathcal{W}_1$ and $\mathcal{W}_2$ satisfy 
\begin{eqnarray}
|\mathcal{V}|\leq |\mathcal{X}|+3, \quad |\mathcal{W}_1|\leq
 |\mathcal{X}|(|\mathcal{X}|+3)+1, \quad |\mathcal{W}_2|\leq
 |\mathcal{X}|(|\mathcal{X}|+3)+1.
\end{eqnarray}
\end{enumerate}

The last two conditions can be removed without causing essential
difference to the region $\mathcal{R}_{in}(D_1,D_2)$; with them
removed, no specific structure is required on the joint distribution
of $(X,V,W_1,W_2)$. To see the last two conditions indeed do not cause
loss of generality, apply the support lemma \cite{Csiszarkorner} as
follows. For an arbitrary joint distribution of $(X,V,W_1,W_2)$
satisfying the first three conditions, we first reduce the cardinality
of $\mathcal{V}$. To preserve $P_X$ and the two distortions and two
mutual information values, $|\mathcal{X}|+3$ letters are needed. With
this reduced alphabet, observe that both the distortion and rate
expressions depend only on the marginal of $(X,V,W_1)$ and
$(X,V,W_2)$, respectively, hence requiring $W_1\leftrightarrow (X,V)
\leftrightarrow W_2$ being a Markov string does not cause any loss of
generality. Next to reduce the cardinality of $\mathcal{W}_1$, it is
seen $|\mathcal{X}||\mathcal{V}|-1$ letters are needed to preserve the
joint distribution of $(X,V)$, one more is needed to preserve $D_1$
and another is needed to preserve $I(X;W_1|Y_1,V)$. Thus
$|\mathcal{X}|(|\mathcal{X}|+3)+1$ letters suffice. Note that we do
not need to preserve the value of $D_2$ and the value of the other
mutual information term because of the aforementioned Markov string. A
similar argument holds for $|\mathcal{W}_2|$.

The following theorem asserts that $\mathcal{R}_{in}(D_1,D_2)$ is an achievable region. 
\begin{theorem}
\label{theorem:achievable}
For any discrete memoryless stochastic source with side informations
under the Markov condition $X\leftrightarrow Y_1 \leftrightarrow Y_2$,
\begin{eqnarray*}
\mathcal{R}(D_1,D_2)\supseteq \mathcal{R}_{in}(D_1,D_2).
\end{eqnarray*}
\end{theorem}

This theorem is proved in Appendix \ref{append:theoremachievable}, and
here we outline the coding scheme for this achievable region in an
intuitive manner.  The encoder first encodes using a $\vec{V}$ codebook with a
``coarse'' binning, such that decoder one is able to decode it with
side information $\vec{Y_1}$. A Wyner-Ziv successive refinement coding
(with side information $\vec{Y_1}$) is then added conditioned on the
codeword $\vec{V}$ also for decoder one using $\vec{W_1}$. The encoder
then enumerates the binning of $\vec{V}$ up to a level such that
$\vec{V}$ is decodable by decoder two using the weaker side
information $\vec{Y_2}$. By doing so, decoder two is able to reduce
the number of possible codewords in the (coarse) bin to a smaller
number, which essentially forms a ``finer" bin; with the weaker side
information $\vec{Y_2}$, the $\vec{V}$ codeword is then decoded
correctly with high probability.  Another Wyner-Ziv successive
refinement coding (with side information $\vec{Y_2}$) is finally added
conditioned on the codeword $\vec{V}$ for decoder two using a random codebook of $\vec{W_2}$.

\begin{figure}[tb]
  \centering 
\includegraphics[scale=0.7]{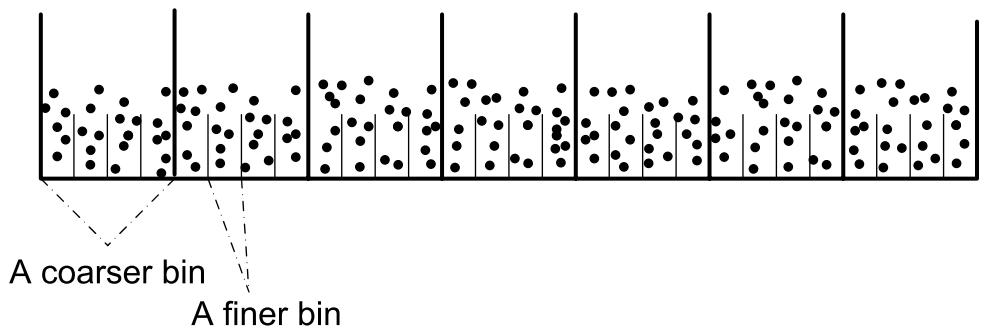}
\caption{An illustration of the codewords in the nested binning
structure. \label{fig:bins}}
\end{figure}

As seen in the above argument, in order to reduce the number of possible $\vec{V}$ codewords from the first stage to the second stage, the key idea is to construct a nested
binning structure as illustrated in Fig. \ref{fig:bins}. Note that this is a fundamentally different from the code structure in SR-WZ, where no nested binning is needed.  
Each of the coarser bin contains the same number of finer bins; each finer bin
holds certain number of codewords. They are constructed in such a way
that given the specific coarser bin index, the first stage decoder can
decode in it with the strong side information; at the second stage,
additional bitstream is received by the decoder, which further
specifies one of the finer bin in the coarser bin, such that the
second stage decoder can decode in this finer bin using the weaker
side information. If we assign each codeword to a finer bin
independently, then its coarser bin index is also independent of that
of the other codewords. 

We note that the coding scheme does not explicitly require that side
informations are degraded. Indeed as long as the chosen random
variable $V$ satisfies $I(V;Y_1)\geq I(V;Y_2)$ as well as the Markov
condition, the region is indeed achievable. More precisely, the following corollary is straightforward.

\begin{corollary}
For any discrete memoryless stochastically source with side
informations $Y_1$ and $Y_2$ (without the Markov structure),
$\widetilde{\mathcal{R}}_{in}(D_1,D_2)\subseteq \mathcal{R}(D_1,D_2)$,
where $\widetilde{\mathcal{R}}_{in}(D_1,D_2)$ is
$\mathcal{R}_{in}(D_1,D_2)$ with the additional condition that
$I(V;Y_1)\geq I(V;Y_2)$.
\end{corollary}

We can specialize the region $\mathcal{R}_{in}(D_1,D_2)$ to give another inner bound. Let $\hat{\mathcal{R}}_{in}(D_1,D_2)$ be the set of all
rate pairs $(R_1,R_2)$ for which there exist random variables
$(W_1,W_2)$ in finite alphabets
$\mathcal{W}_1,\mathcal{W}_2$ such that the following
condition are satisfied.

\begin{enumerate}
\item $W_1\leftrightarrow W_2\leftrightarrow X \leftrightarrow Y_1
\leftrightarrow Y_2$ or $W_2\leftrightarrow W_1\leftrightarrow X \leftrightarrow Y_1
\leftrightarrow Y_2$ is a Markov string.
\item There exist deterministic maps 
$f_j: \mathcal{W}_j\times \mathcal{Y}_j \rightarrow \hat{\mathcal{X}_j}$
such that
\begin{eqnarray}
\Expt\mathit{d}_j(X,f_j(W_j,Y_j))\leq D_j, \quad j=1,2.
\end{eqnarray}
\item The non-negative rate pairs satisfy:
\begin{eqnarray}
R_1\geq I(X;W_1|Y_1),\quad
R_1+R_2\geq I(X;W_2|Y_2)+I(X;W_1|Y_1,W_2).
\end{eqnarray} 
\item The alphabets $\mathcal{W}_1$ and $\mathcal{W}_2$ satisfy 
\begin{eqnarray}
|\mathcal{W}_1|\leq (|\mathcal{X}|+3)(|\mathcal{X}|(|\mathcal{X}|+3)+1), \quad |\mathcal{W}_1|\leq (|\mathcal{X}|+3)(|\mathcal{X}|(|\mathcal{X}|+3)+1).
\end{eqnarray}

\end{enumerate}

\begin{corollary}
\label{cor:inner}
For any discrete memoryless stochastically source with side informations under the Markov condition $X\leftrightarrow Y_1 \leftrightarrow Y_2$,
\begin{eqnarray*}
\mathcal{R}_{in}(D_1,D_2) \supseteq \hat{\mathcal{R}}_{in}(D_1,D_2).
\end{eqnarray*}
\end{corollary}

The region $\hat{\mathcal{R}}_{in}(D_1,D_2)$ is particular interesting for the following reasons. Firstly, it can be explicitly matched back to the coding scheme for the simple Gaussian example. Secondly, it will be shown that one of the outer bounds has the same rate and distortion expressions as $\hat{\mathcal{R}}_{in}(D_1,D_2)$, only with a relaxed Markov string requirement. We now prove this corollary. 

\vspace{0.5cm}
\noindent{\em Proof of Corollary \ref{cor:inner}}

When $W_1\leftrightarrow W_2 \leftrightarrow X$, let $V=W_1$. Then the rate expressions in Theorem \ref{theorem:achievable} gives 
\begin{eqnarray}
R_1\geq I(X;W_1|Y_1),\quad R_1+R_2\geq I(X;V,W_2|Y_2)+I(X;W_1|V,Y_1)=I(X;W_2|Y_2),
\end{eqnarray}
and therefore $\mathcal{R}_{in}(D_1,D_2) \supseteq \hat{\mathcal{R}}_{in}(D_1,D_2)$ for this case. When $W_2\leftrightarrow W_1 \leftrightarrow X$, let $V=W_2$. 
Then the rate expressions in Theorem \ref{theorem:achievable} gives 
\begin{eqnarray*}
R_1&\geq& I(X;V,W_1|Y_1)=I(X;W_1|Y_1)\\ 
R_1+R_2&\geq& I(X;V,W_2|Y_2)+I(X;W_1|V,Y_1)=I(X;W_2|Y_2)+I(X;W_1|W_2,Y_1),
\end{eqnarray*}
and therefore $\mathcal{R}_{in}(D_1,D_2) \supseteq \hat{\mathcal{R}}_{in}(D_1,D_2)$ for this case.

The cardinality bound here is larger than that in Theorem \ref{theorem:achievable} because of the requirement to preserve the Markov conditions. 
\hfill\QED

\subsection{Two outer bounds}
\label{subsec:OuterBnd}

Define the following two regions, which will be shown to be two outer
bounds. An obvious outer bound is given by the intersection of the
Wyner-Ziv rate distortion function and the rate-distortion function
for the problem considered by Heegard and Berger
\cite{HeegardBerger:85} with degraded side information
$X\leftrightarrow Y_1 \leftrightarrow Y_2$
\begin{eqnarray}
\label{eq:IntOtrBnd}
\mathcal{R}_{\cap}(D_1,D_2)=\{(R_1,R_2):R_1\geq
R^*_{X|Y_1}(D_1),\quad R_1+R_2\geq R_{HB}(D_1,D_2)\}.
\end{eqnarray}

A tighter outer bound is now given as follows: define the region
$\mathcal{R}_{out}(D_1,D_2)$ to be the set of all rate pairs
$(R_1,R_2)$ for which there exist random variables $(W_1,W_2)$ in
finite alphabets $\mathcal{W}_1,\mathcal{W}_2$ such that the following
conditions are satisfied.

\begin{enumerate}\label{enm:conditions2}
\item $(W_1,W_2)\leftrightarrow X \leftrightarrow Y_1 \leftrightarrow Y_2$.
\item There exist deterministic maps
$f_j: \mathcal{W}_j\times \mathcal{Y}_j \rightarrow \hat{\mathcal{X}_j}$
such that
\begin{eqnarray}
\Expt\mathit{d}_j(X,f_j(W_j,Y_j))\leq D_j, \quad j=1,2.
\end{eqnarray}
\item $|\mathcal{W}_1|\leq |\mathcal{X}|(|\mathcal{X}|+3)+2$, $|\mathcal{W}_2|\leq |\mathcal{X}|+3$.
\item The non-negative rate vectors satisfies:
\begin{eqnarray}
\label{eqn:rates2}
R_1\geq I(X;W_1|Y_1), \quad
R_1+R_2\geq I(X;W_2|Y_2)+I(X;W_1|Y_1,W_2).
\end{eqnarray} 
\end{enumerate}

The main result of this subsection is the following theorem. 
\begin{theorem}
\label{theorem:outer}
For any discrete memoryless stochastically source with side
informations under the Markov condition $X\leftrightarrow Y_1
\leftrightarrow Y_2$,
\begin{eqnarray*}
\mathcal{R}_{\cap}(D_1,D_2)\supseteq\mathcal{R}_{out}(D_1,D_2)
\supseteq\mathcal{R}(D_1,D_2).
\end{eqnarray*}
\end{theorem}

The first inclusion of
$\mathcal{R}_{\cap}(D_1,D_2)\supseteq\mathcal{R}_{out}(D_1,D_2)$ is
obvious, since $\mathcal{R}_{out}(D_1,D_2)$ takes the same form as
$R^*_{X|Y_1}(D_1)$ and $R_{HB}(D_1,D_2)$ when the rates $R_1$ and
$R_1+R_2$ are considered individually. 
Thus we will focus on the
latter inclusion, whose proof is given in Appendix \ref{appendix:theorem3}. 

Note that the inner bound $\hat{\mathcal{R}}_{in}(D_1,D_2)$ and
$\mathcal{R}_{out}(D_1,D_2)$ have the same rate and distortion
expressions and they differ only by a Markov string requirement
(ignoring the non-essential cardinality bounds). Because of the difference in 
the domain of optimizations, the two bounds may not produce
the same rate-regions. This is quite similar to the case of
distributed lossy source coding problem, for which the Berger-Tung
inner bound requires a long Markov string and the Berger-Tung outer
bound requires only two short Markov strings
\cite{Berger:lecturenotes}, but their rate and distortion expressions
are the same.

\subsection{Lossless reconstruction at one decoder}

Since decoder one has better quality side information, it is
reasonable for it to require a higher quality
reconstruction. Alternatively, from the point of view of universal
coding, when the encoder does not know the quality of the side
information, it might assume the better quality one exists at the
decoder and aim to reconstruct with a higher quality, comparing with
the case when the poorer quality side information is available. In the
extreme case, decoder one might require a lossless reconstruction. In
this subsection, we consider the setting where either decoder one or
decoder two requires lossless reconstruction. We have the following
theorem.

\begin{theorem}
\label{theorem:special}
If $D_1=0$ with $d_1(\cdot,\cdot)\in \Gamma_d$, or $D_2=0$ with
$d_2(\cdot,\cdot)\in\Gamma_d$ (see \ref{eq:GammaDef} for $\Gamma_d$),
then $\mathcal{R}(D_1,D_2)= \mathcal{R}_{in}(D_1,D_2)$. More
precisely, for the former case,
\begin{eqnarray}
\label{eqn:lossless1}
\mathcal{R}(0,D_2)=\bigcup_{P_{W_2}(D_2)}\{(R_1,R_2):R_1\geq H(X|Y_1),
\quad R_1+R_2\geq I(X;W_2|Y_2)+H(X|Y_1,W_2).\},
\end{eqnarray}
where $P_{W_1}(D_2)$ is the set of random variables satisfying the
Markov string $W_2\leftrightarrow X\leftrightarrow Y_1\leftrightarrow
Y_2$, and having a deterministic function $f_2$ satisfying $\Expt
d(f_2(W_2,Y_2),X)\leq D_2$. For the latter case,
\begin{eqnarray}
\label{eqn:lossless2}
\mathcal{R}(D_1,0)=\bigcup_{P_{W_1}(D_1)}\{(R_1,R_2):R_1\geq
I(X;W_1|Y_1), \quad R_1+R_2\geq H(X|Y_2)\},
\end{eqnarray}
where $P_{W_1}(D_1)$ is the set of random variables satisfying the
Markov string $W_1\leftrightarrow X\leftrightarrow Y_1\leftrightarrow
Y_2$, and having a deterministic function $f_1$ satisfying $\Expt
d(f_1(W_1,Y_1),X)\leq D_1$.

\end{theorem}

\vspace{0.5cm}

\noindent{\em Proof of Theorem \ref{theorem:special}:}
For $D_1=0$, let $W_1=X$ and $V=W_2$. The achievable rate vector implied by Theorem \ref{theorem:achievable} is given by 
\begin{eqnarray}
R_1\geq H(X|Y_1), \quad R_1+R_2\geq I(X;W_2|Y_2)+H(X|Y_1,W_2).
\end{eqnarray}
It is seen that this rate region is tight by the converse of Slepian-Wolf coding for rate $R_1$, and by (\ref{eqn:HBcoro}) of Heegard-Berger coding for rate $R_1+R_2$.

For $D_2=0$, let $W_1=V$ and $W_2=X$. The achievable rate vector implied by Theorem \ref{theorem:achievable} is given by 
\begin{eqnarray}
R_1\geq I(X;W_1|Y_1), \quad R_1+R_2\geq H(X|Y_2).
\end{eqnarray}
It is easily seen that this rate region is tight by the converse of
Wyner-Ziv coding for rate $R_1$, and the converse of Slepian-Wolf
coding (or more precisely, Wyner-Ziv rate distortion function
$R_{X|Y_2}(0)$ with $d_2(\cdot,\cdot)\in\Gamma_d$ as given in
\cite{WynerZiv:76}) for rate $R_1+R_2$. \hfill\QED

Zero distortion under a distortion measure $d\in \Gamma_d$ can be
interpreted as {\em lossless}, however, it is a weaker requirement
than that the block error probability is arbitrarily
small. Nevertheless, $\mathcal{R}(0,D_2)$ and $\mathcal{R}(D_1,0)$ in
(\ref{eqn:lossless1}) and (\ref{eqn:lossless2}) still provide valid
outer bounds for the more stringent lossless definition. On the other
hand, it is rather straightforward to specialize the coding scheme for
these cases, and show that the same conclusion is true for lossless
coding in the this case. Thus we have the following corollary.

\begin{corollary}
The rate region, when the first stage, and respectively the second stage,
requires lossless in terms of arbitrary small block error probability
is given by (\ref{eqn:lossless1}), respectively (\ref{eqn:lossless2}),
\end{corollary}

The key difference from the general case when both stages are lossy is
the elimination of the need to generate one of codebooks using an
auxiliary random variables, which simplifies the matter
tremendously. For example when $D_2=0$, since the first stage encoder
guarantees that $\vec{w}_1$ and $\vec{x}$ are jointly typical, the
second stage only needs to construct a codebook of $\vec{x}$ by
binning the approximately $2^{H(X|W_1)}$ such $\vec{x}$ vector
directly. Subsequently the second stage encoder does not search for a
vector $\vec{x}^*$ to be jointly typical with both $\vec{w}_1$ and
$\vec{x}$, but instead just sends the bin index of the observed source
vector $\vec{x}$ directly. Alternatively, it can be understood as both
the encoder and decoder at the second stage have access to a side
information vector $\vec{w}_1$, and thus a conditional Slepian-Wolf
coding with decoder side information $Y_2$ suffices.

\subsection{Deterministic distortion measure}

Another case of interest is when some functions of the source $X$ is
required to be reconstructed with arbitrary small distortion in terms
of Hamming distortion; see \cite{Gam:82} for the corresponding case
for the multiple description problem. More precisely, let $Q_i:
\mathcal{X}\rightarrow \mathcal{Z}_i$, $i=1,2$ be two deterministic
functions and denote $Z_i=Q_i(X)$.  Consider the case that decoder $i$
seeks to reconstruct $Z_i$ with arbitrarily small Hamming distortion
\footnote{By a similar argument as in the last subsection, the same
result holds if block error probability is made arbitrarily
small.}. The achievable region $\mathcal{R}_{in}$ is tight when the
functions satisfy certain degradedness condition as stated in the
following theorem.

\begin{theorem}
\label{theorem:deterministic}
Let the distortion measure be Hamming distortion $d_H:
\mathcal{Z}_i\times \mathcal{Z}_i\rightarrow \{0,1\}$ for $i=1,2$.
\begin{enumerate}
\item If there exists a deterministic function
$Q':\mathcal{Z}_1\rightarrow \mathcal{Z}_2$ such that $Q_2=Q'\cdot
Q_1$, then $\mathcal{R}(0,0)= \mathcal{R}_{in}(0,0)$. More precisely
\begin{eqnarray}
\label{eq:RatesDetDist1}
\mathcal{R}(0,0)=\left\{(R_1,R_2): R_1\geq H(Z_1|Y_1),\,R_1+R_2\geq
H(Z_2|Y_2)+H(Z_1|Y_1Z_2)\right\}.
\end{eqnarray}
\item If there exists a deterministic function
$Q':\mathcal{Z}_2\rightarrow \mathcal{Z}_1$ such that $Q_1=Q'\cdot
Q_2$, then $\mathcal{R}(0,0)= \mathcal{R}_{in}(0,0)$. More precisely
\begin{eqnarray}
\label{eq:RatesDetDist2}
\mathcal{R}(0,0)=\left\{(R_1,R_2): R_1\geq H(Z_1|Y_1),\,R_1+R_2\geq
H(Z_2|Y_2)\right\}.
\end{eqnarray}
\end{enumerate}
\end{theorem}
\vspace{0.5cm}

\noindent{\em Proof of Theorem \ref{theorem:deterministic}:} To prove
(\ref{eq:RatesDetDist1}), first observe that by letting $W_1=Z_1$ and
$V=W_2=Z_2$, $\mathcal{R}_{in}$ clearly reduces to the given
expression. For the converse, we start from the outer bound
$\mathcal{R}_{out}(0,0)$, which implies that $Z_1$ is a function of
$W_1$ and $Y_1$, and $Z_2$ is a function of $W_2$ and $Y_2$. For the
first stage rate $R_1$, we have the following chain of equalities
\begin{eqnarray}
R_1\geq I(X;W_1|Y_1)=I(X;W_1Z_1|Y_1)\geq I(X;Z_1|Y_1)=H(Z_1|Y_1)-H(Z_1|X,Y_1)=H(Z_1|Y_1).
\end{eqnarray}
For the sum rate, we have 
\begin{eqnarray*}
R_1+R_2&\geq& I(X;W_2|Y_2)+I(X;W_1|W_2Y_1)\\
&=&I(X;W_2Z_2|Y_2)+I(X;W_1|W_2Y_1)\\
&=&I(X;Z_2|Y_2)+I(X;W_2|Y_2Z_2)+I(X;W_1|W_2Y_1)\\
&=&H(Z_2|Y_2)+I(X;W_2|Y_2Z_2)+I(X;W_1|W_2Y_1)\\
&\stackrel{(a)}{\geq}&H(Z_2|Y_2)+I(X;W_2|Y_1Y_2Z_2)+I(X;W_1|W_2Y_1)\\
&\stackrel{(b)}{=}&H(Z_2|Y_2)+I(X;W_2|Y_1Y_2Z_2)+I(X;W_1|W_2Y_1Y_2)\\
&=&H(Z_2|Y_2)+I(X;W_2|Y_1Y_2Z_2)+I(X;W_1|W_2Y_1Y_2Z_2)\\
&=&H(Z_2|Y_2)+I(X;W_1W_2|Y_1Y_2Z_2)\\
&\geq&H(Z_2|Y_2)+I(X;Z_1|Y_1Y_2Z_2)\\
&=&H(Z_2|Y_2)+H(Z_1|Y_1Y_2Z_2)\\
&\stackrel{(c)}{=}&H(Z_2|Y_2)+H(Z_1|Y_1Z_2),
\end{eqnarray*}
where (a) is due to the Markov string $W_2\leftrightarrow X \leftrightarrow (Y_1Y_2)$ and $Z_2$ is function of $X$; (b) is due to the Markov string $(W_1W_2)\leftrightarrow X \leftrightarrow Y_1 \leftrightarrow Y_2$; (c) is due to the Markov string $(Z_1,Z_2)\leftrightarrow Y_1\leftrightarrow Y_2$.

Proof of part 2) ({\em i.e.,} (\ref{eq:RatesDetDist2}) relationship)
is straightforward and is omitted.  \hfill\QED

Clearly in the converse proof, the requirement that the functions
$Q_1$ and $Q_2$ are degraded is not needed. Indeed this outer bound
holds for any general functions, however the degradedness is needed
for establishing the achievability of the region. If the coding is not
necessarily scalable, then it can be seen the sum rate is indeed achievable, and the result above can be used to establish a non-trivial special result in the context of the problem treated by
Heegard and Berger \cite{HeegardBerger:85}.
\begin{corollary}
Let the two function $Q_1$ and $Q_2$ be arbitrary, and let the distortion measure be Hamming distortion $d_H: \mathcal{Z}_i\times \mathcal{Z}_i\rightarrow \{0,1\}$ for $i=1,2$, then we have 
\begin{eqnarray}
R_{HB}(0,0)=H(Z_2|Y_2)+H(Z_1|Y_1Z_2).
\end{eqnarray}

\end{corollary}

\section{Perfect Scalability and a Binary Source}
\label{sec:perfScalable}

In this section we introduce the notion of perfect scalability, which
is defined as when both the stages operate at the Wyner-Ziv rates. We
further examine the doubly symmetric binary source and provide a
partial characterization and investigate its scalability. The
quadratic Gaussian source with jointly Gaussian side informations is
treated in Section \ref{sec:Gaussian} in a more general setting.

\subsection{Perfect Scalability}

The notion of the (strict) successive refinability defined in
\cite{SteinbergMerhav:04} for the SR-WZ problem with forward
degradation in the side-informations (SI) can be applied to the
reversely degraded case considered in this paper. This is done by
introducing the notion of perfect scalability for the SI-scalable
problem defined below.


\begin{definition}
A source $X$ is said to be {\em perfectly scalable} for distortion pair
$(D_1,D_2)$, with side informations under the Markov string $X
\leftrightarrow Y_1 \leftrightarrow Y_{2}$, if
\begin{eqnarray}
(R^*_{X|Y_1}(D_1),R^*_{X|Y_2}(D_2)-R^*_{X|Y_1}(D_1))\in
\mathcal{R}(D_1,D_2).\nonumber
\end{eqnarray}
\end{definition}


\begin{theorem}
\label{theorem:perfect}
A source $X$ with side informations under the Markov string $X
\leftrightarrow Y_1 \leftrightarrow Y_{2}$, for which $\exists \,\, y_1\in \mathcal{Y}_1$ such that
$P_{XY_1}(x,y_1)>0$ for each $x\in \mathcal{X}$, is perfectly scalable
for distortion pair $(D_1,D_2)$ if and only if there exist random
variables $(W_1,W_2)$ and deterministic maps
$f_j:\mathcal{W}_j\times\mathcal{Y}_j\rightarrow\hat{\mathcal{X}}_j$
such that the following conditions hold simultaneously:
\begin{enumerate}
\item $R_{X|Y_j}^*(D_j)=I(X;W_j|Y_j)$ and $\Expt
d_j(X,f_j(W_1,Y_j))\leq D_j$, for $j=1,2$.
\item $W_1\leftrightarrow W_2\leftrightarrow X\leftrightarrow Y_1
\leftrightarrow Y_2$ forms a Markov string.
\item The alphabet $\mathcal{W}_1$ and $\mathcal{W}_2$ satisfy 
$|\mathcal{W}_1|\leq |\mathcal{X}|(|\mathcal{X}|+3)+2$, and
$|\mathcal{W}_2|\leq |\mathcal{X}|+3$.
\end{enumerate}
\end{theorem}

The Markov string is the most crucial condition, and the substring
$W_1\leftrightarrow W_2\leftrightarrow X$ is the same as one of the
condition for successive refinability without side information
\cite{EquitzCover:91}\cite{Rimoldi:94}. The support condition
essentially requires the existence of a worst letter $y_1$ in the
alphabet $\mathcal{Y}_1$ such that it has non-zero probability mass
for each $(x,y_1)$ pair, $x\in \mathcal{X}$.

\vspace{0.3cm}
\noindent{\em Proof of Theorem \ref{theorem:perfect}}

The sufficiency being trivial, we only prove the necessity.  Without
loss of generality, assume $P_X(x)>0$ for all $x\in \mathcal{X}$.  By
Theorem \ref{theorem:outer}, if
$(R^*_{X|Y_1}(D_1),R^*_{X|Y_2}(D_2)-R^*_{X|Y_1}(D_1)$ is achievable
for $(D_1,D_2)$, then using the tighter outer bound
$\mathcal{R}_{out}(D_1,D_2)$ of Theorem \ref{theorem:outer}, there
exist random variable $W_1,W_2$ in finite alphabet, whose sizes is
bounded as $|\mathcal{W}_1|\leq |\mathcal{X}|(|\mathcal{X}|+3)+2$ and
$|\mathcal{W}_2|\leq |\mathcal{X}|+3$, and functions $f_1,f_2$ such
that $(W_1,W_2)\leftrightarrow X \leftrightarrow Y_1 \leftrightarrow
Y_2$ is a Markov string, $\Expt d_j(X,f_j(W_j,Y_j))\leq D_j$ for
$j=1,2$ and
\begin{eqnarray}
R_{X|Y_1}^*(D_1)\geq I(X;W_1|Y_1),\quad R_{X|Y_2}^*(D_2)\geq
I(X;W_2|Y_2)+I(X;W_1|Y_1,W_2).
\end{eqnarray}

It follows 
\begin{eqnarray}
\label{eqn:nec}
R_{X|Y_2}^*(D_2)\geq I(X;W_2|Y_2)+I(X;W_1|Y_1,W_2)
\geq  I(X;W_2|Y_2)\stackrel{(a)}{\geq} R_{X|Y_2}^*(D_2),
\end{eqnarray}
where (a) follows the converse of rate-distortion theorem for
Wyner-Ziv coding. Since the leftmost and the rightmost quantities are
the same, all the inequalities must be equalities in (\ref{eqn:nec}),
and it follows $I(X;W_1|Y_1,W_2)=0$. Similarly we have
\begin{eqnarray}
\label{eqn:necR1}
R_{X|Y_1}^*(D_1)\geq I(X;W_1|Y_1)\geq R_{X|Y_1}^*(D_1),
\end{eqnarray}
thus (\ref{eqn:necR1}) also holds with equality. 

Notice that if $W_1\leftrightarrow W_2 \leftrightarrow X$ is a Markov
string, then we can use Corollary \ref{cor:inner} to claim the
sufficiency and complete the proof. However, this Markov condition is
not true in general. This is where the support condition is needed.

For convenience, define the set 
\begin{equation}
\label{eq:Fdef}
F(w_2)=\{x\in
\mathcal{X}:P(x,w_2)>0\}.
\end{equation} 
By the Markov string $(W_1,W_2)\leftrightarrow X \leftrightarrow Y_1$,
the joint distribution of $(w_1,w_2,x,y_1)$ can be factorized as
follows
\begin{eqnarray}
P(w_1,w_2,x,y_1)=P(x,y_1)P(w_2|x)P(w_1|x,w_2).
\end{eqnarray}
Furthermore, $I(X;W_1|Y_1,W_2)=0$ implies the Markov string $X\leftrightarrow (W_2,Y_1)\leftrightarrow W_1$, and thus the joint distribution of $(w_1,w_2,x,y_1)$ can also be factorized as follows
\begin{eqnarray}
P(w_1,w_2,x,y_1)=P(x,y_1,w_2)p(w_1|y_1,w_2)
\stackrel{(a)}{=}P(x,y_1)P(w_2|x)P(w_1|y_1,w_2),
\end{eqnarray}
where (a) follows by the Markov substring $W_2\leftrightarrow X
\leftrightarrow Y_1 \leftrightarrow Y_2$.  Fix an arbitrary
$(w_1^*,w_2^*)$ pair, by the assumption that $P(x,y_1)>0$ for any
$x\in \mathcal{X}$, we have
\begin{eqnarray}
P(w_2^*|x)P(w_1^*|x,w_2^*)=P(w_2^*|x)P(w_1^*|y_1,w_2^*)
\end{eqnarray}
for any $x\in \mathcal{X}$. Thus for any $x\in F(w_2^*)$ (see
definition in (\ref{eq:Fdef})) such that $P(w_1|x,w_2^*)$ is well
defined, we have
\begin{eqnarray}
p(w_1^*|y_1,w_2^*)=p(w_1^*|x,w_2^*)
\end{eqnarray}
and it further implies 
\begin{eqnarray}
p(w_1^*|w_2^*)=\frac{\sum_x P(x,w_1^*,w_2^*)}{\sum_x
P(x,w_2^*)}=\frac{\sum_{x\in F(w_2^*)} P(x,w_2^*)P(w_1^*|y_1,w_2^*)}{\sum_x
P(x,w_2^*)}=p(w_1^*|y_1,w_2^*)=p(w_1^*|x,w_2^*)
\end{eqnarray}
for any $x\in F(w_2^*)$. This indeed implies $W_1\leftrightarrow W_2 \leftrightarrow X$ is a Markov string, which completes the proof.
\hfill\QED

\subsection{The Doubly Symmetric Binary Source with Hamming Distortion Measure}
\label{subsection:DSBS}
Consider the following source: $X$ is a memoryless binary source
$X\in\{0,1\}$ and $P(X=0)=0.5$. The first stage side information $Y$
can be taken as the output of a binary symmetric channel with input
$X$, and crossover probability $p<0.5$. The second stage does not have
side information. This source clearly satisfies the support condition
in Theorem \ref{theorem:perfect}. It will be shown that for some
distortion pairs, this source is perfectly scalable, while for others
this is not possible. We next first provide partial results using
$\hat{\mathcal{R}}_{in}$ and $\mathcal{R}_{\cap}$ previously given.

\begin{figure}[tb]
  \centering \includegraphics[scale=0.5]{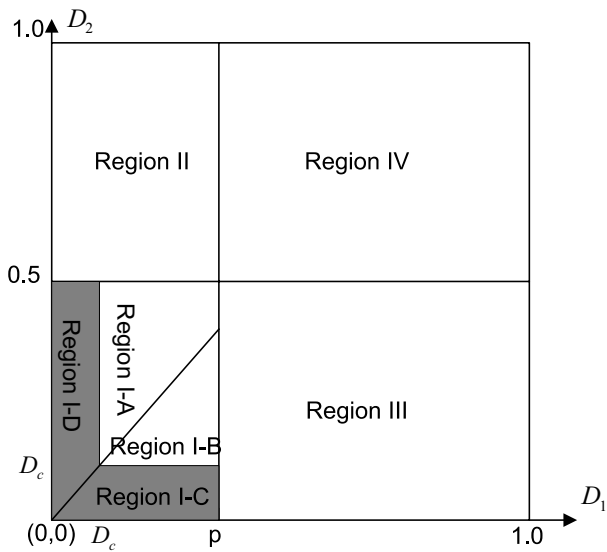}
\caption{The partition of the distortion region, where $d_c$ is the
critical distortion in \cite{WynerZiv:76} below which time sharing is
not necessary. \label{fig:regions}}
\end{figure}

An explicit calculation of $R_{HB}(D_1,D_2)$, together with the
optimal forward test channel structure, was given in a recent work
\cite{TianDiggavi:05}.  With this explicit calculation, it can be
shown that in the shaded region in Fig. \ref{fig:regions}, the outer
bound $\mathcal{R}_{\cap}(D_1,D_2)$ is in fact achievable (as well as
in Region II, III and IV; however these three regions are degenerate
cases, and will be ignored in what follows). Recall the definition of
the critical distortion $d_c$ in the Wyner-Ziv problem for the DSBS
source in \cite{WynerZiv:76}
\begin{eqnarray*}
\frac{G(d_c)}{d_c-p}=G'(d_c),
\end{eqnarray*}
where $G(u)=h_b(p*u)-h_b(u)$, $h_b(u)$ is the binary entropy function
$h_b(u)=-u \log u-(1-u)\log (1-u)$, and $u*v$ is the binary
convolution for $0\leq u,v\leq 1$ as $u*v=u(1-v)+v(1-u)$.  It was
shown in \cite{WynerZiv:76} that if $D\leq d_c$, then
$R^*_{X|Y}(D)=G(D)$. We will use the following result from
\cite{TianDiggavi:05}.
\begin{theorem}
For distortion pairs $(D_1,D_2)$ such that $0\leq D_2\leq0.5$ and
$0\leq D_1\leq\min(d_c,D_2)$ ({\em i.e.,} Region I-D),
\begin{eqnarray*}
R_{HB}(D_1,D_2)=1-h_b(D_2*p)+G(D_1).
\end{eqnarray*}
\end{theorem}

\begin{figure}[tb]
  \centering
    \includegraphics[width=12cm]{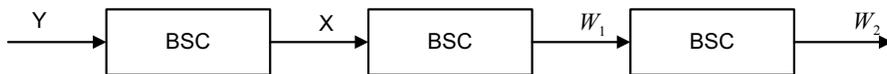}
\caption{The forward test channel in Region
I-D\label{fig:special}. The crossover probability for the BSC between
$X$ and $W_1$ is $D_1$, while the crossover probability $\eta$ for the
BSC between $W_1$ and $W_2$ is such that $D_1*\eta=D_2$.}
\end{figure}

This result implies that for the shaded region I-D, the forward test
channel to achieve this lower bound is in fact a cascade of two BSC
channels depicted in Fig. \ref{fig:special}. This choice clearly
satisfies the condition in Corollary \ref{cor:inner} with the rates
given by the outer bound $\mathcal{R}_{\cap}(D_1,D_2)$, which shows
that this outer bound is indeed achievable. Note the following
inequality
\begin{eqnarray}
R_{HB}(D_1,D_2)=1-h_b(D_2*p)+h_b(p*D_1)-h_b(D_1)\geq 1-h_b(D_2)=R(D_2),
\end{eqnarray}
where the inequality is due to the monotonicity of $G(u)$ in $0\leq
u\leq 0.5$, we conclude that in this regime the source is not
perfectly scalable.

\begin{figure}[tb]
  \centering
    \includegraphics[width=12cm]{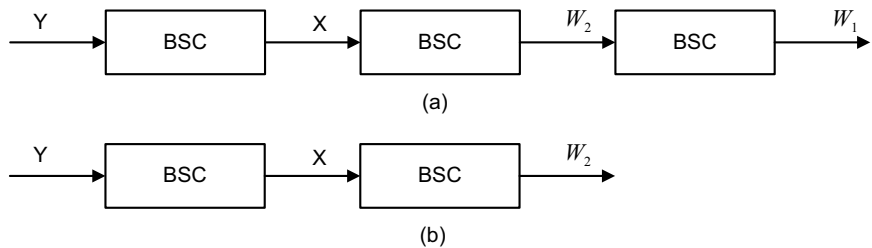}
\caption{The forward test channels in Region
I-C. The crossover probability for the BSC between
$X$ and $W_2$ is $D_2$ in both the channels, while the crossover
probability $\eta$ for the BSC between $W_2$ and $W_1$ in (a) is such
that $D_2\leq D_1*\eta=\eta'\leq d_c$. Note for (b), $W_1$ can be
taken as a constant.}
\label{fig:specialC}
\end{figure}

To see $\mathcal{R}_{\cap}(D_1,D_2)$ is also achievable in region I-C,
recall the result in \cite{WynerZiv:76} that the optimal forward test
channel to achieve $R^*_{X|Y}(D)$ has the following structure: it is
the time-sharing between zero-rate coding and a BSC with crossover
probability $d_c$ if $D\geq d_c$, or a single BSC with crossover
probability $D$ otherwise. Thus it is straightforward to verify that
$\mathcal{R}_{\cap}(D_1,D_2)$ is achievable by time sharing the two
forward test channels in Fig. \ref{fig:specialC}; furthermore, an
equivalent forward test channel can be found such that the Markov
condition $W_1'\leftrightarrow W_2\leftrightarrow X$ is satisfied,
which satisfies the conditions given in Theorem
\ref{theorem:perfect}. Thus in this regime, the source is in fact
perfectly scalable.

\begin{figure}[tb]
  \centering \includegraphics[width=12cm]{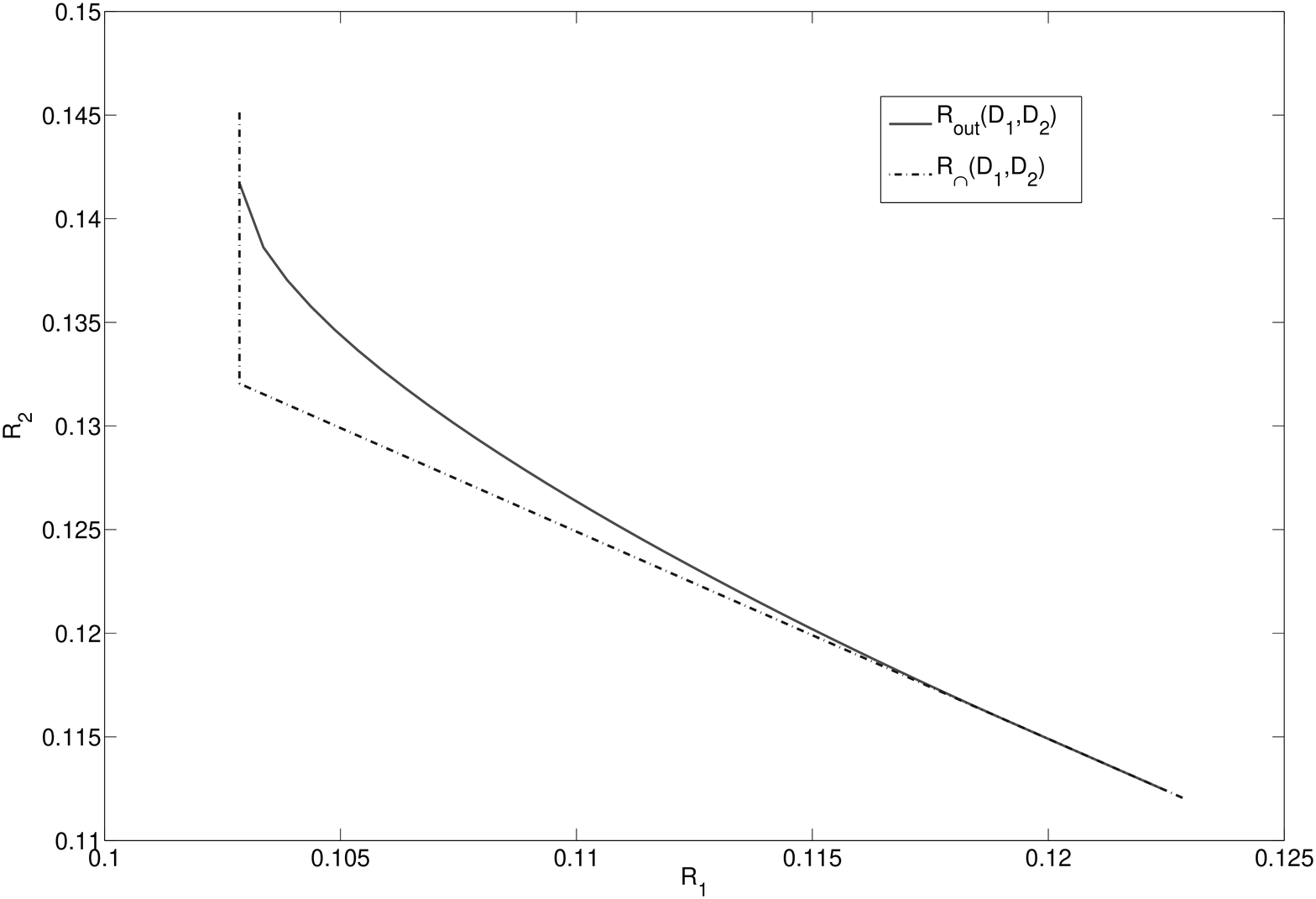}
\caption{The rate outer bounds for a particular choice of $D_1,D_2$ in
Region I-B of Figure \ref{fig:regions}. \label{fig:DSBS}}
\end{figure}

Unfortunately, we were not able to find the complete characterization
for the regime I-A and I-B. Using an approach similar to
\cite{TianDiggavi:05}, an explicit outer bound can be derived from
$\mathcal{R}_{out}(D_1,D_2)$.  It can then be shown numerically that
for certain distortion pairs in this regime,
$\mathcal{R}_{out}(D_1,D_2)$ is strictly tighter than
$\mathcal{R}_{\cap}(D_1,D_2)$. This calculation can be found in
\cite{TianDiggavi:report06} and is omitted here. An example is given in
Fig. \ref{fig:DSBS} for the two outer bounds with a non-zero gap in
between for a specific distortion pair in Region I-B.


\section{A Near Sufficiency Result}
\label{sec:rateloss}
  
\begin{figure}[tb]
  \centering
    \includegraphics[width=9cm]{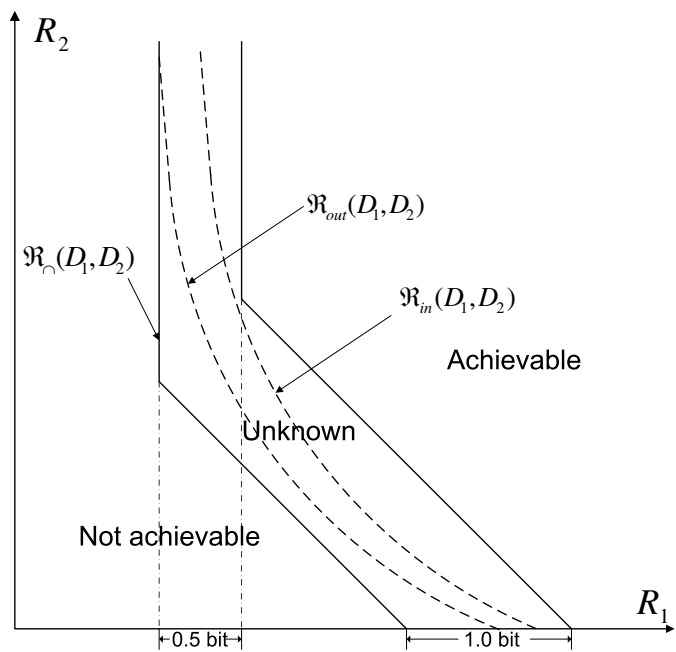}
\caption{An illustration of the gap between the inner bound and the
outer bounds when MSE is the distortion measure. The two regions
$\mathcal{R}_{in}(D_1,D_2)$ and $\mathcal{R}_{out}(D_1,D_2)$ are given
in dashed lines, since it is unknown whether they are indeed the
same.\label{fig:rateloss}}
\end{figure}

By using the tool of rate loss introduced by Zamir \cite{Zamir:96A},
which was further developed in \cite{Lastras:01,Lastras:06,Feng:03,Feng:06},
it can be shown that when both the source and reconstruction alphabets
are reals, and the distortion measure is MSE, the gap between the
achievable region and the out bounds are bounded by a constant. Thus the inner and outer bounds are nearly sufficient in the sense defined in \cite{Lastras:06}. 
To show this result, we distinguish the two cases $D_1\geq D_2$ and $D_1\leq
D_2$. The source $X$ is assumed to have finite variance
$\sigma_x^2$ and finite (differential) entropy. The result of this section is summarized in
Fig. \ref{fig:rateloss}.

\subsection{The case $D_1\geq D_2$} 

Construct two random variable $W_1'=X+N_1+N_2$ and $W_2'=X+N_2$, where
$N_1$ and $N_2$ are zero mean independent Gaussian random variables,
independent of everything else, with variance $\sigma_1^2$ and
$\sigma_2^2$ such that $\sigma_1^2+\sigma_2^2=D_1$ and
$\sigma_2^2=D_2$. By letting $V'=W'_1$, it is obvious that
the following rates are achievable for distortion $(D_1,D_2)$ from
Theorem \ref{theorem:achievable}
\begin{eqnarray}
\label{eq:NSr1r2}
R_1=I(X;X+N_1+N_2|Y_1),\quad R_1+R_2=I(X;X+N_2|Y_2).
\end{eqnarray} 

Let $U$ be optimal random variable to achieve the
Wyner-Ziv rate at distortion $D_1$ given decoder side information
$Y_1$. Then it is clear that the difference between $R_1$ and the Wyner-Ziv rate can be bounded as,\begin{eqnarray}
&&I(X;X+N_1+N_2|Y_1)-I(X;U|Y_1)\nonumber\\
&\stackrel{(a)}{=}&I(X;X+N_1+N_2|UY_1)-I(X;U|Y_1,X+N_1+N_2)\nonumber\\
&\leq&I(X;X+N_1+N_2|UY_1)\nonumber\\
&=&I(X-\hat{X}_1;X-\hat{X}_1+N_1+N_2|UY_1)\nonumber\\ &\leq&
I(X-\hat{X}_1,U,Y_1;X-\hat{X}_1+N_1+N_2)\nonumber\\
&=&I(X-\hat{X}_1;X-\hat{X}_1+N_1+N_2)+I(U,Y_1;X-\hat{X}_1+N_1+N_2|X-\hat{X}_1)\nonumber\\
&=&I(X-\hat{X}_1;X-\hat{X}_1+N_1+N_2)\nonumber\\
&\stackrel{(b)}{\leq}&\frac{1}{2}\log_2\frac{D_1+D_1}{D_1}=0.5
\label{eqn:rateloss}
\end{eqnarray}
where $(a)$ is by applying chain rule to $I(X;X+N_1+N_2,U|Y_1)$ in two
different ways; $(b)$ is true because $\hat{X}_1$ is the decoding
function given $(U,Y_1)$, the distortion between $X$ and $\hat{X}_1$
is bounded by $D_1$, and $X-\hat{X}_1$ is independent of $(N_1,N_2)$.

Now we turn to bound the gap for the sum rate $R_1+R_2$. Let $W_1$ and $W_2$ be the two random variables to achieve the rate distortion function $R_{HB}(D_1,D_2)$. First notice
the following two identities due to the Markov string
$(W_1,W_2)\leftrightarrow X \leftrightarrow Y_1\leftrightarrow Y_2$
and $(N_1,N_2)$ are independent of $(X,Y_1,Y_2)$
\begin{eqnarray}
I(X;W_2|Y_2)+I(X;W_1|W_2Y_1)&=&I(X;W_1W_2|Y_1)+I(Y_1;W_2|Y_2)\\
I(X;X+N_2|Y_2)&=&I(X;X+N_2|Y_1)+I(Y_1;X+N_2|Y_2).
\end{eqnarray}
Next we can bound the difference between the sum-rate $R_1+R_2$ (as given in (\ref{eq:NSr1r2}))
and the Heegard-Berger sum rate as follows. 
\begin{eqnarray}
&&I(X;X+N_2|Y_2)-I(X;W_2|Y_2)-I(X;W_1|W_2Y_1)\nonumber\\
&=&\{I(X;X+N_2|Y_1)-I(X;W_1W_2|Y_1)\}+\{I(Y_1;X+N_2|Y_2)-I(Y_1;W_2|Y_2)\}.
\end{eqnarray}
To bound the first bracket, notice that 
\begin{eqnarray}
&&I(X;X+N_2|Y_1)-I(X;W_1W_2|Y_1)\nonumber\\
&=&I(X;X+N_2|W_1W_2Y_1)-I(X;W_1W_2|Y_1,X+N_2)\nonumber\\
&\leq&I(X;X+N_2|W_1W_2Y_1)\nonumber\\
&\stackrel{(a)}{=}&I(X;X+N_2|W_1W_2Y_1Y_2)\nonumber\\
&=&I(X-\hat{X}_2;X-\hat{X}_2+N_2|W_1W_2Y_1Y_2)\nonumber\\
&\leq&I(X-\hat{X}_2,W_1,W_2,Y_1,Y_2;X-\hat{X}_2+N_2)\nonumber\\
&=&I(X-\hat{X}_2;X-\hat{X}_2+N_2)+I(W_1,W_2,Y_1,Y_2;X-\hat{X}_2+N_2|X-\hat{X}_2)\nonumber\\
&=&I(X-\hat{X}_2;X-\hat{X}_2+N_2)\leq
\frac{1}{2}\log_2\frac{D_2+D_2}{D_2}=0.5
\end{eqnarray}
where (a) is due to the Markov string $(W_1,W_2)\leftrightarrow
X\leftrightarrow Y_1\leftrightarrow Y_2$, $\hat{X}_2$ is the decoding
function given $(W_2,Y_2)$, and the other inequalities follow similar
arguments as in Eqn. (\ref{eqn:rateloss}). To bound the second
bracket, we write the following
\begin{eqnarray}
I(Y_1;X+N_2|Y_2)-I(Y_1;W_2|Y_2)\nonumber\\
&=&I(Y_1;X+N_2|W_2Y_2)-I(Y_1;W_2|Y_2,X+N_2)\nonumber\\
&\leq&I(Y_1;X+N_2|W_2Y_2)\nonumber\\
&\leq&I(XY_1;X+N_2|W_2Y_2)\nonumber\\
&=&I(X;X+N_2|W_2Y_2)\leq \frac{1}{2}\log_2\frac{D_2+D_2}{D_2}=0.5
\label{eqn:secondbracket}
\end{eqnarray}

Thus we have shown that for $D_1\geq D_2$, the gap between the outer
bound $\mathcal{R}_{\cap}(D_1,D_2)$ and the inner bound
$\mathcal{R}_{in}(D_1,D_2)$ is bounded. More precisely, the gap for
$R_1$ is bounded by 0.5 bit, while the gap for the sum rate is bounded
by 1.0 bit.

\subsection{The case $D_1\leq D_2$} 

Construct random variable $W_1'=X+N_1$ and $W_2'=X+N_1+N_2$, where
$N_1$ and $N_2$ are zero mean independent Gaussian random variables,
independent of everything else, with variance $\sigma_1^2$ and
$\sigma_2^2$ such that $\sigma_1^2=D_1$ and
$\sigma_1^2+\sigma_2^2=D_2$. By letting $V'=W_2'=X+N_1+N_2$, it is easily seen that the following
rates are achievable for distortion $(D_1,D_2)$
\begin{eqnarray*}
R_1&=&I(X;X+N_1|Y_1)\\
R_1+R_2&=&I(X;X+N_1+N_2|Y_2)+I(X;X+N_1|Y_1,X+N_1+N_2).
\end{eqnarray*}

Clearly, the argument for the first stage $R_1$ still holds with minor
changes. To bound the sum-rate gap, notice the following identity
\begin{eqnarray}
&&I(X;X+N_1+N_2|Y_2)+I(X;X+N_1|Y_1,X+N_1+N_2)\nonumber\\
&=&I(X;X+N_1+N_2|Y_1)+I(Y_1;X+N_1+N_2|Y_2)+I(X;X+N_1|Y_1,X+N_1+N_2)\\
&=&I(Y_1;X+N_1+N_2|Y_2)+I(X;X+N_1|Y_1).
\end{eqnarray}
Next we seek to upper bound the following quantity
\begin{eqnarray}
&&I(X;X+N_1+N_2|Y_2)+I(X;X+N_1|Y_1,X+N_1+N_2)-I(X;W_2|Y_2)-I(X;W_1|W_2Y_1)\nonumber\\
&&=\{I(X;X+N_1|Y_1)-I(X;W_1W_2|Y_1)\}+\{I(Y_1;X+N_1+N_2|Y_2)-I(Y_1;W_2|Y_2)\},
\end{eqnarray}
where again $W_1,W_2$ are the R-D optimal random variables for $R_{HB}(D_1,D_2)$. 
For the first bracket, we have 
\begin{eqnarray}
&&I(X;X+N_1|Y_1)-I(X;W_1W_2|Y_1)\nonumber\\
&=&I(X;X+N_1|W_1W_2Y_1)-I(X;W_1W_2|Y_1,X+N_1)\nonumber\\
&\leq&I(X;X+N_1|W_1W_2Y_1)\nonumber\\
&=&I(X-\hat{X}_1;X-\hat{X}_1+N_2|W_1W_2Y_1)\nonumber\\
&\leq&I(X-\hat{X}_1,W_1,W_2,Y_1;X-\hat{X}_1+N_2)\nonumber\\
&=&I(X-\hat{X}_1;X-\hat{X}_1+N_1)+I(W_1,W_2,Y_1;X-\hat{X}_1+N_1|X-\hat{X}_1)\nonumber\\
&=&I(X-\hat{X}_1;X-\hat{X}_1+N_1)\nonumber\\
&\leq& \frac{1}{2}\log\frac{D_1+D_1}{D_1}=0.5,
\end{eqnarray}
where $\hat{X}_1$ is the decoding function given $(W_1,Y_1)$. For the
second bracket, following a similar approach as
(\ref{eqn:secondbracket}), we have
\begin{eqnarray*}
&&I(Y_1;X+N_1+N_2|Y_2)-I(Y_1;W_2|Y_2)\\
&\leq& I(X;X+N_1+N_2|W_2Y_2)\\
&\leq&I(X-\hat{X}_2,W_2,Y_2;X-\hat{X}_2+N_1+N_2)\\
&=&I(X-\hat{X}_2;X-\hat{X}_2+N_1+N_2)\leq 0.5
\end{eqnarray*}
Thus we conclude that for both cases the gap between the inner bound
and the outer bound is bounded. Fig. \ref{fig:rateloss} illustrates
the inner bound and outer bounds, as well as the gap in between.

\section{The Quadratic Gaussian Source with Jointly Gaussian Side Informations}
\label{sec:Gaussian}


The degraded side information assumption, either $X\leftrightarrow
Y_1\leftrightarrow Y_2$ or $X\leftrightarrow Y_2\leftrightarrow Y_1$,
for the quadratic jointly Gaussian case is especially interesting,
since physically degradedness and stochastic degradedness
\cite{CoverThomas} do not cause essential difference in terms of the
rate-distortion region for the problem being considered \cite{SteinbergMerhav:04}. Moreover, jointly
Gaussian source-side information is always statistically degraded,
these forwardly and reversely degraded cases together provide a
complete solution to the jointly Gaussian case with two decoders.

In this section we in fact consider a more general setting with an
arbitrary number of decoders for jointly Gaussian source and multiple
side informations. Though the source and side informations can have
arbitrary correlation, in light of the discussion above, we will treat
only physically degraded side informations. Note that since a specific
encoding order is specified, though the side informations are degraded
as an unordered set, the quality of side informations may not be
monotonic along the scalable coding order. Clearly the solution for
the two stage case can be reduced in a straightforward manner from the
general solution. Recall from Theorem \ref{theorem:outer} (see (\ref{eq:IntOtrBnd})) that $\mathcal{R}_{\cap}(D_1,D_2)$ is an outer 
bound derived from the intersection of the Heegard-Berger and
Wyner-Ziv bounds. The generalization of the outer bound
$\mathcal{R}_{\cap}(D_1,D_2)$ to $N$ decoders plays an important role,
and therefore we take a detour in Section \ref{subsec:GaussianHB} to
start with the characterization of $R_{HB}(D_1,D_2,\ldots  ,D_N)$ for the
jointly Gaussian case.

\subsection{$R_{HB}(D_1,D_2,\ldots  ,D_N)$ for the jointly Gaussian case}
\label{subsec:GaussianHB}

Consider the following source $X \sim \mathcal{N}(0,\sigma_x^2)$, and
side informations $Y_k=X+\sum_{i=1}^k N_i$, where $N_i \sim
\mathcal{N}(0,\sigma_i^2)$ are mutually independent and independent of
$X$. The result by Heegard and Berger \cite{HeegardBerger:85} gives
\begin{eqnarray}
\label{eq:NdecHB}
R_{HB}(D_1,D_2,\ldots  ,D_N)=\min_{p(D_1,D_2,\ldots  ,D_N)}\sum_{k=1}^N
I(X;W_k|Y_k,W_{k+1},W_{k+2},\ldots  ,W_N),
\end{eqnarray}
where $p(D_1,D_2,\ldots  ,D_N)$ is the set of all random variable with the
Markov string $(W_1,W_2,\ldots  ,W_N)\leftrightarrow X \leftrightarrow
(Y_1,Y_2,\ldots  ,Y_N)$, such that deterministic functions
$f_k(Y_k,W_k,W_{k+1},\ldots ,W_N)$, $k=1,\ldots  ,N$ exist which satisfy the
distortion constraints.  In \cite{TianDiggavi:05}, the case $N=2$ was
calculated explicitly, however such an explicit calculation appears
quite involved for general $N$ due to the discussion of various cases
when some of the distortion constraints are not tight. In the
sequel we approach the problem by showing a jointly Gaussian forward
test channel is optimal.

Note that if we choose to enforce only a subset of the distortion
constraints, the rate for such a restriction gives a lower bound on
$R_{HB}(D_1,D_2,\ldots  ,D_N)$. By taking all the non-empty subsets of the
distortion constraints, labeled by elements of $I_N=\{1,2,\ldots  ,N\}$, a
total of $2^N-1$ lower bounds are available and clearly the maximum of
them is also a lower bound. More precisely, we are interested in $\max
R^*_{HB}(A_D)$, where $A_D\subseteq I_N$ and $R^*_{HB}(A_D)$ is defined
in the sequel explicitly in terms of the distortion constraints only;
note that if $i\in A_D$, $D_i$ is still the distortion constraint for
the decoder with side information $Y_i$. We next derive one of these
lower bounds using all the constraints $(D_1,D_2,\ldots  ,D_N)$,
i.e. $A_D=I_N$; a similar derivation applies to the case with any
subset $A_D\subset I_N$. Using (\ref{eq:NdecHB}) we have,
\begin{eqnarray*}
&&\sum_{k=1}^N I(X;W_k|Y_k,W_{k+1},W_{k+2},\ldots  ,W_N)\\
&=&h(X|Y_N)-h(X|Y_1W_1^N)-h(X|Y_NW_N)+h(X|Y_{N-1}W_N)\\
&&\qquad\qquad-h(X|Y_{N-1}W_{N-1}^N)+
\ldots +h(X|Y_1W_2^N)\nonumber\\
&\stackrel{(a)}{=}&h(X|Y_N)-h(X|Y_1W_1^N)\\
&&\qquad\qquad-[h(X|Y_NW_N)-h(X|Y_{N-1}Y_NW_N)]-
\ldots -[h(X|Y_2W_2^N)-h(X|Y_1Y_2W_2^N)]\nonumber\\
&=&h(X|Y_N)-h(X|Y_1W_1^N)-I(X;Y_{N-1}|Y_NW_N)\\
&&\qquad\qquad-I(X;Y_{N-2}|Y_{N-1}W_{N-1}^N)-\ldots -I(X;Y_1|Y_2W_2^N)\\
&\stackrel{(b)}{=}&h(X|Y_N)-h(X|Y_1W_1^N)\\
&&\qquad\qquad-[h(Y_{N-1}|Y_NW_N)-h(Y_{N-1}|XY_N)]-\ldots -[h(Y_1|Y_2W_2^N)-h(Y_1|Y_2X)]\\
&=&h(X|Y_N)+\sum_{k=2}^N
h(Y_{k-1}|XY_k)-\sum_{k=2}^Nh(Y_{k-1}|Y_kW_k^N)-h(X|Y_1,W_1^N),
\end{eqnarray*}
where (a) is because of the Markov string $X\leftrightarrow
(Y_{k-1}W_k^N)\leftrightarrow Y_{k}$, and (b) is because of the Markov
string $W_k^N \leftrightarrow (XY_{k}) \leftrightarrow Y_{k-1}$, both
of which are consequences of $W_k^N\leftrightarrow
X\leftrightarrow Y_{k-1}\leftrightarrow Y_k$. 
The first two terms depend only on the
source and distribution $P_{XY_1\ldots Y_N}$, and we now seek to bound
the latter two terms, for which we have
\begin{eqnarray}
\label{eqn:HX}
h(X|Y_1W_1^N)=h(X-\Expt(X|YW_1^N)|YW_1^N)\leq h(X-\Expt(X|YW_1^N))\leq
h(\mathcal{N}(0,D_1))=\frac{1}{2}\log(2\pi e D_1),
\end{eqnarray}
where the second inequality is because Gaussian distribution maximizes
the entropy for a given second moment, and
$\Expt(X-\Expt(X|YW_1^N))^2\leq D_1$ by the existence of the decoding
function $f_1$. Next define
\begin{eqnarray}
\gamma_k=\frac{\sum_{i=1}^{k-1}\sigma_i^2}{\sum_{i=1}^{k}\sigma_i^2},\, k=2,3,...,N.
\end{eqnarray}
and write the following 
\begin{eqnarray}
Y_{k-1}&=&X+\sum_{i=1}^{k-1}N_i=X+\sum_{i=1}^{k-1}N_i+\gamma_k\sum_{i=1}^k
N_i-\gamma_k \sum_{i=1}^k N_i\\
&=&\gamma_k(X+\sum_{i=1}^{k}N_i)+(1-\gamma_k)X+[\sum_{i=1}^{k-1}N_i-\gamma_k\sum_{i=1}^k
N_i]\\
&=&\gamma_kY_k+(1-\gamma_k)X+[\sum_{i=1}^{k-1}N_i-\gamma_k\sum_{i=1}^k
N_i]
\end{eqnarray}
Notice that 
\begin{eqnarray}
\Expt[Y_k(\sum_{i=1}^{k-1}N_i-\gamma_k\sum_{i=1}^k
N_i)]=\sum_{i=1}^{k-1}\sigma_i^2-\gamma_k\sum_{i=1}^k\sigma_i^2=0,
\end{eqnarray}
and $Y_k$ and $(\sum_{i=1}^{k-1}N_i-\gamma_i\sum_{i=1}^k N_i)$ are
jointly Gaussian, which implies that they are independent. Furthermore
because $(\sum_{i=1}^{k-1}N_i-\gamma_i\sum_{i=1}^k N_i)$ is
independent of $X$, the Markov string $(Y_1,Y_2,\ldots Y_N)\leftrightarrow
X \leftrightarrow (W_1,W_2,\ldots ,W_N)$ implies that it is also
independent of $(W_1,W_2,\ldots ,W_N)$. It follows
\begin{eqnarray}
h(Y_{k-1}|Y_kW_k^N)&=&h\left(\gamma_kY_k+(1-\gamma_k)X+\sum_{i=1}^{k-1}N_i-\gamma_k\sum_{i=1}^k N_i|Y_kW_k^N\right)\\
&=&h\left((1-\gamma_k)X+\sum_{i=1}^{k-1}N_i-\gamma_k\sum_{i=1}^k N_i|Y_kW_k^N\right)\\
&=&h\left((1-\gamma_k)(X-\Expt(X|Y_kW_k^N))+\sum_{i=1}^{k-1}N_i-\gamma_k\sum_{i=1}^k N_i|Y_kW_k^N\right)\\
&\leq&h\left((1-\gamma_k)(X-\Expt(X|Y_kW_k^N))+\sum_{i=1}^{k-1}N_i-\gamma_k\sum_{i=1}^k N_i\right).
\label{eqn:HYk}
\end{eqnarray}
By the aforementioned independence relation, the variance of term in the bracket is bounded above by 
\begin{eqnarray}
\hat{D}_k\stackrel{\Delta}{=}(1-\gamma_k)^2D_k+(1-\gamma_k)^2\sum_{i=1}^{k-1}\sigma_i^2+\gamma_k^2\sigma_k^2.
\end{eqnarray}
Define the following quantities
\begin{eqnarray}
K_1&\stackrel{\Delta}{=}&h(X|Y_N)=\frac{1}{2}\log \frac{2\pi e
\sigma_x^4}{\sigma_x^2+\sum_{i=1}^N\sigma_i^2},\\
K_k&\stackrel{\Delta}{=}&h(Y_{k-1}|XY_k)=\frac{1}{2}\log \frac{2\pi e
\sigma_k^4}{\sum_{i=1}^k\sigma_i^2}, \quad k=2,3,\ldots ,N
\end{eqnarray}
Summarizing the bounds in (\ref{eqn:HX}) and (\ref{eqn:HYk}), we have
\begin{eqnarray}
R_{HB}(D_1,D_2,\ldots D_N)\geq\frac{1}{2}\log\frac{\prod_{i=1}^N
K_i}{\prod_{i=1}^N \hat{D}_i}\stackrel{\Delta}{=}R_{HB}^*(I_N),
\end{eqnarray}
where for convenience we define $\hat{D}_1=D_1$. 

To show that $\max_{A_D\subseteq\{D_1,D_2,\ldots ,D_N\}} R^*_{HB}(A_D)$ is
indeed achievable, construct the random variables
$(W_1^*,W_2^*,\ldots ,W_N^*)$ as follows. Assume that $D_k\leq
\Expt[X-\Expt(X|Y_k)]^2$ for each $k=1,2,\ldots ,N$, because otherwise
this distortion requirement can be ignored completely.

\noindent\textbf{[Construction of $(W_1^*,W_2^*,\ldots ,W_N^*)$]}
\begin{enumerate}
\item For each $k=1,2,\ldots ,N$, determine the variance $\sigma_{Z_k}^2$
of a Gaussian random variable $Z_k$ such that
$D_k=\Expt[X-\Expt(X|Y_k,X+Z_k)]^2$.
\item Rank the variance of $\sigma_{Z_k}^2$ in an increasing order,
and let $\omega(k)$ denote the rank of $\sigma_{Z_k}^2$.
\item Calculate $\sigma^2_{Z'_{1}}=\sigma^2_{Z_{\omega^{-1}(1)}}$, and
$\sigma^2_{Z'_{k}}=\sigma^2_{Z_{\omega^{-1}(k)}}-\sigma^2_{Z_{\omega^{-1}(k-1)}}$
for $k=2,3,\ldots ,N$.
\item Construct a set of independent zero-mean Gaussian random
variables $(Z'_1,Z'_2,\ldots ,Z'_N)$ to have variance $\sigma^2_{Z'_{k}}$.
\item Construct a set of random variables $(W_1^*,W_2^*,\ldots ,W_N^*)$ as 
\begin{eqnarray}
W_k^*=X+\sum_{i=1}^{\omega(k)} Z'_{k}.
\end{eqnarray}
\end{enumerate}

Next we show that this construction of $(W_1^*,W_2^*,\ldots ,W_N^*)$ achieves
one of aforementioned lower bounds and thus is an optimal forward test
channel. Choose the set $A^*_D=\{k:\omega(k)< \omega(j)\,\,\mbox{for
all}\, j>k\}$, and denote the rank (in increasing order) of its
element $k$ as $r(k)$.  Clearly by the construction we have
\begin{eqnarray*}
&&\sum_{k=1}^N I(X;W^*_k|Y_k,W^*_{k+1},W^*_{k+2},\ldots ,W^*_N)\\
&=&\sum_{k\in A^*_D} I(X;W^*_k|Y_k,W^*_{k+1},W^*_{k+2},\ldots ,W^*_N)\\
&=&\sum_{j=1}^{|A^*_D|}I(X;W^*_{r^{-1}(j)}|Y_{r^{-1}(j)},W^*_{r^{-1}(j+1)})\label{eqn:rateAD}\\
&=&h(X|Y_{r^{-1}(|A^*_D|)})-h(X|W^*_{r^{-1}(|A^*_D|)}Y_{r^{-1}(|A^*_D|)})\\
&&+h(X|Y_{r^{-1}(|A^*_D|-1)}W^*_{r^{-1}(|A^*_D|)})-h(X|Y_{r^{-1}(|A^*_D|-1)}W^*_{r^{-1}(|A^*_D|-1)})\\&&+\ldots +h(X|Y_{r^{-1}(1)}W^*_{r^{-1}(2})-h(X|Y_{r^{-1}(1)}W^*_{r^{-1}(1)})\\
&=&h(X|Y_{r^{-1}(|A^*_D|)})-h(X|Y_{r^{-1}(1)}W^*_{r^{-1}(1)})\\
&&-[h(Y_{r^{-1}(|A^*_D|-1)}|Y_{r^{-1}(|A^*_D|)}W^*_{r^{-1}(|A^*_D|)})-h(Y_{r^{-1}(|A^*_D|-1)}|XY_{r^{-1}(|A^*_D|)})]\\
&&-\ldots -[h(Y_{r^{-1}(1)}|Y_{r^{-1}(2)}W^*_{r^{-1}(2)})-h(Y_{r^{-1}(1)}|XY_{r^{-1}(2)})]\\
&=&R_{HB}^*(A^*_D)
\end{eqnarray*} 
because of the construction of $(W_1^*,W_2^*,\ldots ,W_N^*)$ and the fact that they are jointly Gaussian with $(X,Y_1,Y_2,\ldots ,Y_N)$. Thus, we have proved the following theorem.

\begin{theorem}
The auxiliary random variable $(W_1^*,W_2^*,\ldots ,W_N^*)$ constructed above achieves the minimum in the Heegard and Berger rate distortion function for the jointly Gaussian source and side informations.
\end{theorem}

It is clear that we can determine the set $A^*_D$ before constructing
$(W^*_1,W^*_2,\ldots ,W^*_N)$ using the aforementioned procedure, which can
simplify the construction. However, the current construction has the
advantage that each $W^*_k$ is almost individually determined by $D_k$,
and does not substantially depend on the other distortion
constraints. This will prove to be useful for the general scalable coding
problem. It is worth noting that it seemingly requires comparing
$2^N-1$ values of $R_{HB}^*(A_D)$ to determine
$R_{HB}(D_1,D_2,\ldots ,D_2)$, however, from the forward calculation we
see that in fact $O(N)$ complexity suffices.

\begin{figure}[tb]
  \centering
    \includegraphics[width=10cm]{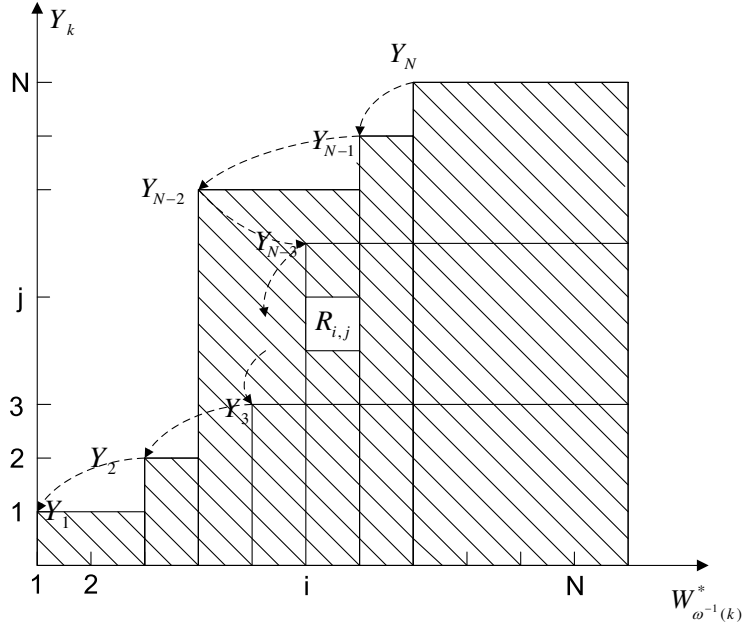}
\caption{An illustration of the sum-rate for the Gaussian case.\label{fig:rate}} 
\end{figure}

This result can be interpreted using Fig. \ref{fig:rate}. On the
horizontal axis, the $N$ marks stand for the $N$ random variable
$(W^*_{\omega^{-1}(1)},W^*_{\omega^{-1}(2)},\ldots ,W^*_{\omega^{-1}(N)})$,
and the on the vertical axis, the $N$ marks stand for the $N$ levels
of side informations $(Y_1,Y_2,\ldots ,Y_N)$. The random variable pairs $(W_k,Y_k)$ are then the points of interest on the plane, since if the $k$-th decoder has $(Y_k,W_k)$ the desired distortion can be achieved; the $(W_k,Y_k)$ pairs are in one-to-one correspondence to the $(\omega(k),k)$ pairs. 
Next we associate the unit square below and to the right of each integer point $(i,j)$ is associated with a rate of value 
\begin{eqnarray}
\label{eqn:Rij}
R_{i,j}=I(W_{\omega^{-1}(i)};Y_{j-1}|Y_{j}W_{\omega^{-1}(i+1)})
\end{eqnarray}
where we define $W_{\omega^{-1}(N+1)}=\emptyset$, and $Y_{0}=X$. For
each $k=1,2,\ldots ,N$, if we cover the rectangle below and to the right
of $(\omega(k),k)$, then the sum rate associated with the covered area
is exactly $R_{HB}(D_1,D_2,\ldots ,D_N)$.

With Fig. \ref{fig:rate}, the coding scheme can be understood as
follows. The coding proceeds from $Y_N$ to $Y_1$, {\em i.e.,} from
high to low on the vertical axis; the $k$-th step ($k$-th decoder) specifies an integer point $(\omega(k),k)$, which corresponds to a $(W_k,Y_k)$ pair, 
on the figure, and additional rate is required if the area below and
to the right of this point induces new area to cover. This order is
illustrated in Fig. \ref{fig:rate} along the arrows. Note that
\begin{eqnarray}
\sum_{j=1}^k R_{i,j}&=&\sum_{j=1}^k
I(W_{\omega^{-1}(i)};Y_{j-1}|Y_{j}W_{\omega^{-1}(i+1)})\\
&=&\sum_{j=1}^k
[I(W_{\omega^{-1}(i)};Y_{j-1}|W_{\omega^{-1}(i+1)})-I(W_{\omega^{-1}(i)};Y_{j}|W_{\omega^{-1}(i+1)})]\\
&=&I(W_{\omega^{-1}(i)};X|W_{\omega^{-1}(i+1)})-I(W_{\omega^{-1}(i)};Y_k|W_{\omega^{-1}(i+1)})]\\
&=&I(W_{\omega^{-1}(i)};X|Y_kW_{\omega^{-1}(i+1)}),
\end{eqnarray}
and it is the rate for a vertical slice of hight $k$ between
horizontal position $i$ and $i+1$, which is in a quite similar form as
(\ref{eqn:rateAD}). In this example figure, the decoders with side
information $Y_{N-3}$ and $Y_3$ do not require additional rates. More
generally, if $(\omega(k),k)$ is inside the area already covered by
the previous coding steps $(N,N-1,\ldots ,k+1)$, then this stage does not
require additional rates. In fact, the corners of the final covered area
specifies the set $A^*_D$.

The following observations are essential for the general Gaussian
scalable coding problem: each unit square in Fig. \ref{fig:rate} is
not merely associated with rate $R_{i,j}$, it is in fact associated
with a fraction of code $C_{i,j}$ with the following properties
\begin{enumerate}
\item The rate of $C_{i,j}$ is (asymptotically) $R_{i,j}$;
\item If the fractions of code associated with the area below and to
the right of $(\omega(k),k)$ are available, then the decoder with side
information $Y_k$ can decode within distortion $D_k$;
\item The same set of code $C_{i,j}$ can be used to fulfill only
subset of the constraints, the rate calculated by the covering area
method is the quadratic Gaussian Heegard and Berger rate distortion
function.
\end{enumerate}
The first and second observations are straightforward by constructing
the nested binning together with conditional codebooks as described in
Section \ref{sec:Ach}, {\em i.e.,} $N-1$ conditioning stage from
$W^*_{\omega^{-1}(1)}$ to $W^*_{\omega^{-1}(N)}$ and each conditioned
codebook has $N$ nested levels from coarse for $Y_1$ to fine for $Y_N$. In fact, it is not necessary to use $N$ nested level for each codebook, but we do so for simplicity of
understanding. The last property is due to the inherent Markov string
among $W^*_1,W^*_2,\ldots ,W^*_N$ and $X$.

\subsection{Scalable coding with joint Gaussian side informations}

\begin{figure}[tb]
  \centering
    \includegraphics[width=10cm]{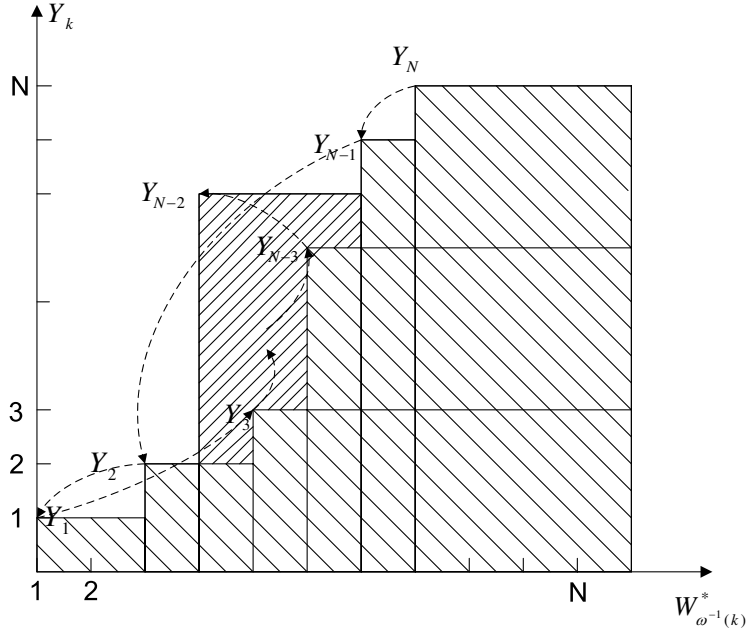}
\caption{An illustration of incremental rate for scalable
coding.\label{fig:rate2} The denser shaded region gives the
incremental rate $R_k$ for the stage with side information $Y_k$.}
\end{figure}

Now consider the scalable coding problem where side informations and
distortions are given by a permutation $\pi(\cdot)$ of that in the
last subsection, {\em i.e.,} $Y'_i=Y_{\pi(i)}$ and
$D'_i=D_{\pi(i)}$. We next show that the identically permuted set of
random variable $(W_1^*,W_2^*,\ldots ,W_N^*)$ achieves the Heegard-Berger rate
distortion function for any first $k$ stages, thus optimal. In light
of pictorial interpretation in Fig. \ref{fig:rate}, this reduces to
rearranging the coded stream of $C_{i,j}$. Fig. \ref{fig:rate2} shows
the effect of changing the scalable coding order.

More precisely, for a certain side information $Y'_i=Y_{\pi(i)}$,
define the following sets:
\begin{eqnarray}
C(k)&=&\{\pi(i):i<k,\pi(i)>\pi(k)\}\\
E_{-}(k)&=&\{\pi(i):i<k,\pi(i)<\pi(k),\omega(\pi(i))>\omega(\pi(k))\},
\end{eqnarray} 
and the following function
\begin{eqnarray}
E(k)=\max \left[\{\pi(i):i<k,\pi(i)<\pi(k),\omega(\pi(i))<\omega(\pi(k))\}\cup\{0\}\right], 
\end{eqnarray}
and let $Y_0=X$. 
Let the set of integers $E_-(k)$ be ordered increasingly, and the rank of its element $j$ be $r(j)$.
Denote the set of random variables $\{W_{j}:j\in C\}$ as $W^*_C$ for
an integer set $C$.  The following $k$-th stage rate is achievable for
$k=1,2,\ldots ,N$
\begin{eqnarray*}
\label{eqn:kthR}
R_k&=&\sum_{i=1}^{|E_-(k)|}
I(Y_{r^{-1}(i)};W^*_{r^{-1}(i)}|Y_{\pi(k)}W^*_{r^{-1}(i+1)}W^*_{r^{-1}(i+2)},\ldots ,W^*_{r^{-1}(|E_-(k)|)}W^*_{C(k)})\\
&&+I(Y_{E(k)};W^*_{\pi(k)}|Y_{\pi(k)}W^*_{E_-(k)}W^*_{C(k)}).
\end{eqnarray*}
It is clearly this rate corresponds to exactly the dense shaded region
in Fig. \ref{fig:rate2}, which is the sum of rates of fraction of
codes $C(i,j)$ as described above. The property of this fraction code
$C(i,j)$ thus implies the following.
\begin{theorem}
The Gaussian scalable coding achievable rate region for distortion vector $(D_{\pi(1)},D_{\pi(2)},\ldots ,D_{\pi(N)})$ is the rate vectors $(R_1,R_2,\ldots ,R_N)$ satisfies
\begin{eqnarray}
\sum_{i=1}^k R_i\geq R_{HB}(D_{\pi(1)},D_{\pi(2)},\ldots ,D_{\pi(k)}),\quad k=1,2,...,N
\end{eqnarray}
where the side informations are $(Y_{\pi(1)},Y_{\pi(2)},\ldots ,Y_{\pi(k)})$. Furthermore, it is achievable by a jointly Gaussian codebook with nested binning.
\end{theorem}

An immediate consequence of this result is the following corollary.
\begin{corollary}
A distortion vector $(D_{\pi(1)},D_{\pi(2)},\ldots ,D_{\pi(N)})$ is perfectly scalable along side informations $(Y_{\pi(1)},Y_{\pi(2)},\ldots ,Y_{\pi(k)})$ for the jointly Gaussian source if and only if $R_{HB}(D_{\pi(1)},D_{\pi(2)},\ldots ,D_{\pi(k)})=R^*_{X|Y_{\pi(k)}}(D_{\pi(k)})$ for each $k=1,2,\ldots ,N$.
\end{corollary}

This corollary applies to one of the important special cases where
$D_1=D_2=\ldots =D_N$ and $\pi(k)=N-k+1$ for each $k$, {\em i.e.,} when
all the decoders have the same distortion requirement, and the
scalable order is along a decreasing order of side information
quality. This implies that at least for the Gaussian case, an
opportunistic coding strategy does exist when the distortion
requirement is the same for all the users.

\section{Conclusion}
\label{sec:disc}
We studied the problem of scalable source coding with reversely
degraded side-information and gave two inner bounds as well as two
outer bounds. These bounds are tight for special cases such as one lossless decoder and under certain deterministic distortion measures. 
Furthermore we provided a complete solution to the
Gaussian source with quadratic distortion measure with any number of
jointly Gaussian side informations.  The problem of perfect
scalability is investigated and the gap between the inner and outer
bounds are shown to be bounded. For the doubly symmetric binary source
with Hamming distortion, we provided partial results of the
rate-distortion region. The result illustrates the difference between
the lossless and the lossy source coding: though a universal approach
exists with uncertain side informations at the decoder for the
lossless case, such uncertainty generally causes loss of performance
in the lossy case.

\appendices

\section{Notation and Basic Properties of Typical Sequences}
We will follow the definition of typicality in \cite{Csiszarkorner},
but use a slightly different notation to make the small positive
quantity $\delta$ explicit (see \cite{SteinbergMerhav:04}).

\begin{definition}
A sequence $\vec{x}\in \mathcal{X}^n$ is said to be
$\delta$-strongly-typical with respect to a distribution $P_X(x)$ on
$\mathcal{X}$ if
\begin{enumerate}
\item For all $a\in \mathcal{X}$ with $P_X(a)>0$
\begin{eqnarray}
\left|\frac{1}{n}N(a|\vec{x})-P_X(a)\right|<\delta,
\end{eqnarray}
\item For all $a\in \mathcal{X}$ with $P_X(a)=0$, $N(a|\vec{x})$=0,
\end{enumerate}
where $N(a|\vec{x})$ is the number of occurrences of the symbol $a$ in
the sequence $\vec{x}$. The set of sequences $\vec{x}\in
\mathcal{X}^n$ that is $\delta$-strongly-typical is called the
$\delta$-strongly-typical set and denoted as $T_{[X]}^{\delta}$, where
the dimension $n$ is dropped.
\end{definition}

The following properties are well-known and will be used in the proof:
\begin{enumerate}

\item Given a $\vec{x}\in T_{[X]}^\delta$, for a $\vec{y}$ whose
component is drawn i.i.d according to $P_Y$ and any $\delta'>\delta$,
we have
\begin{eqnarray}
2^{-n(I(X;Y)+\lambda_1)}\leq P[(\vec{x},\vec{y})\in T_{[XY]}^{\delta'}] \leq 2^{-n(I(X;Y)-\lambda_1)}
\end{eqnarray}
where $\lambda_1$ is a small positive quantity $\lambda_1\rightarrow
0$ as $n \rightarrow \infty$ and both $\delta, \delta'\rightarrow 0$.
\item Similarly, given $(\vec{x},\vec{y})\in T_{[XY]}^{\delta'}$, for
any $\delta''>\delta'$, let the component of $\vec{z}$ be drawn i.i.d
according to the conditional marginal $P_{Z_i|Y_i}(y_i)$, then
\begin{eqnarray}
2^{-n(I(X;Z|Y)+\lambda_2)}\leq P[(\vec{x},\vec{y},\vec{z})\in
T_{[XYZ]}^{\delta''}] \leq 2^{-n(I(X;Z|Y)-\lambda_2)}
\end{eqnarray}
where $\lambda_2$ is a small positive quantity $\lambda_2\rightarrow
0$ as $n \rightarrow \infty$ and both $\delta', \delta''\rightarrow
0$.
\item {\em Markov Lemma \cite{Berger:lecturenotes}:} If
$X\leftrightarrow Y \leftrightarrow Z$ is a Markov string, and
$\vec{X}$ and $\vec{Y}$ are such that their component is drawn
independently according to $P_{XY}$. Then for all $\delta>0$
\begin{eqnarray}
\lim_{n\rightarrow \infty}P[(\vec{X},\vec{z})\in
T_{[XZ]}^{|\mathcal{Y}|\delta}\left|\right(\vec{Y},\vec{z})\in
T_{[YZ]}^\delta]\rightarrow 1.
\end{eqnarray}
furthermore,
\begin{eqnarray}
\lim_{n\rightarrow \infty}P[(\vec{X},\vec{Y},\vec{z})\in
T_{[XYZ]}^{\delta}\left|\right(\vec{Y},\vec{z})\in
T_{[YZ]}^\delta]\rightarrow 1.
\end{eqnarray}
\end{enumerate}

\section{Proof of Theorem \ref{theorem:achievable}}
\label{append:theoremachievable}

{\em Codebook generation: } Let a probability distribution
$P_{W_1W_2XY_1Y_2}=P_{XVW_1W_2}P_{Y_1|X}P_{Y_2|Y_1}$, and two
reconstruction functions $f_1(Y_1,W_1)$ and $f_2(Y_2,W_2)$ be
given. First construct $2^{nR_A}$ coarser bins and $2^{nR_A+R_A'}$
finer bins, where $R_A$ and $R_A'$ are to be specified later. Generate
$2^{R_V}$ length-$n$ codewords according to $P_V(\cdot)$, denote this
set of codewords as $\mathcal{C}_v$; assign each of them into one of
the finer bins independently. For each codeword
$\vec{v}\in\mathcal{C}_v$, generate $2^{nR_{W_1}}$ length-$n$
codewords according to
$P_{W_1|V}(\vec{w_1}|\vec{v})=\prod_{k=1}^nP_{W_1|V}(w_{1,k}|v_k)$,
denote this set of codewords as $\mathcal{C}_{W_1}(\vec{v})$;
independently assign each codeword to one of the $2^{nR_B}$
bins. Again for each $\vec{V}$ codeword, independently generate
$2^{nR_{W_2}}$ length-$n$ codewords according to
$P_{W_2|V}(\vec{w_2}|\vec{v})=\prod_{k=1}^nP_{W_2|V}(w_{2,k}|v_k)$,
denote this set of codewords as $\mathcal{C}_{W_2}(\vec{v})$;
independently assign each codeword to one of the $2^{nR_C}$
bins. Reveal this codebook to the encoders and decoders.

{\em Encoding:} For a given $\vec{x}$, find in $\mathcal{C}_v$ a
codeword $\vec{v}^*$ such that $(\vec{x},\vec{v}^*)\in
T_{[XV]}^{2\delta}$; calculate the coarser bin index $i(\vec{v}^*)$,
and the finer bin index within the coarser bin $j(\vec{v}^*)$. Then in the
$\mathcal{C}_{w_1}(\vec{v}^*)$ codebook, find a codeword $\vec{w}_1^*$
such that $(\vec{w}_1^*,\vec{v}^*,\vec{x}^*)\in
T_{[W_1VX]}^{3\delta}$, and calculate its corresponding bin index
$k$. In $\mathcal{C}_{w_2}(\vec{v}^*)$ codebook, find a codeword
$\vec{w}_2^*$ such that $(\vec{w}_2^*,\vec{v}^*,\vec{x})\in
T_{[W_2VX]}^{3\delta}$, and calculate its corresponding bin index
$l$. The first-stage encoder sends $i$ and $k$, and the second-stage
encoder sends $j$ and $l$. In the above procedure, if there is more than
one joint-typical sequence, choose the least; if there is none, choose
a default codeword and declare an error.

{\em Decoding:} The first stage decoder finds $\hat{\vec{v}}$ in the
coarser bin $i$, such that $(\hat{\vec{v}},\vec{y_1})\in
T_{[VY_1]}^{3|\mathcal{X}|\delta}$; then in the
$\mathcal{C}_{w_1}(\hat{\vec{v}})$ codebook, find $\hat{\vec{w_1}}$
such that $(\hat{\vec{w_1}},\hat{\vec{v}},\vec{y_1})\in
T_{[W_1VY_1]}^{4|\mathcal{X}|\delta}$.  In the second stage, the
decoder finds $\hat{\vec{v}}$ in the finer bin specified by $(i,j)$
such that $(\hat{\vec{v}},\vec{y_2})\in
T_{[VY_2]}^{3|\mathcal{X}|\delta}$; then in the
$\mathcal{C}_{w_2}(\hat{\vec{v}})$ codebook, find $\hat{\vec{w_2}}$
such that $(\hat{\vec{w_2}},\hat{\vec{v}},\vec{y_2})\in
T_{[W_2VY_2]}^{4|\mathcal{X}|\delta}$.  In the above procedure, if
there is none or there are more than one, an error is declared and the
decoding stops. The first decoder reconstructs as
$\hat{x}_{1,k}=f_1(\hat{w}_{1,k},y_{1,k})$ and the second decoder as
$\hat{x}_{2,k}=f_2(\hat{w}_{2,k},y_{2,k})$.

{\em Probability of error:} First define the encoding errors:
\begin{eqnarray}
E_0&=&\{\vec{X}\notin T_{[X]}^{\delta}\}\cup \{\vec{Y_1}\notin
T_{[Y_1]}^{\delta}\}\cup \{\vec{Y_2}\notin
T_{[Y_2]}^{\delta}\}\nonumber\\ E_1&=&E_0^c\cap\{\forall
\vec{v}\in\mathcal{C}_v, (\vec{X},\vec{v})\notin
T_{[XV]}^{2\delta}\}\nonumber\\ E_2&=&E_0^c\cap E_1^c\cap\{\forall
\vec{w_1}\in\mathcal{C}_{w_1}(\vec{v}^*),
(\vec{w_1},\vec{v}^*,\vec{X})\notin T_{[W_1VX]}^{3\delta}\}\nonumber\\
E_3&=&E_0^c\cap E_1^c\cap\{\forall
\vec{w_2}\in\mathcal{C}_{w_2}(\vec{v}^*),
(\vec{w_2},\vec{v}^*,\vec{X})\notin T_{[W_2VX]}^{3\delta}\}\nonumber.
\end{eqnarray}
Next define the decoding errors:
\begin{eqnarray}
E_4&=&E_0^c\cap E_1^c\cap\{(\vec{v}^*,\vec{X},\vec{Y_1})\notin T_{[VXY_1]}^{2\delta}\}\nonumber\\
E_5&=&E_0^c\cap E_1^c\cap\{(\vec{v}^*,\vec{X},\vec{Y_2})\notin T_{[VXY_2]}^{2\delta}\}\nonumber\\
E_6&=&E_0^c\cap E_1^c\cap\{\exists \vec{v}'\neq\vec{v}^*:
i(\vec{v}')=i(\vec{v}^*) \ \text{and} \ (\vec{v}',\vec{Y_1})\in
T_{[VY_1]}^{3|\mathcal{X}|\delta}\}\nonumber\\ E_7&=&E_0^c\cap
E_1^c\cap\{\exists \vec{v}'\neq\vec{v}^*: i(\vec{v}')=i(\vec{v}^*)\
\text{and} \ j(\vec{v}')=j(\vec{v}^*) \ \text{and} \
(\vec{v}',\vec{Y_2})\in T_{[VY_2]}^{3|\mathcal{X}|\delta}\}\nonumber\\
E_8&=&E_0^c\cap E_1^c\cap E_2^c\cap E_4^c\cap
E_6^c\cap\{(\vec{w^*_1},\vec{v^*},\vec{X},\vec{Y_1})\notin
T_{[W_1VXY_1]}^{3\delta} \}\nonumber\\ E_9&=&E_0^c\cap E_1^c\cap
E_3^c\cap E_5^c\cap
E_7^c\cap\{(\vec{w^*_2},\vec{v^*},\vec{X},\vec{Y_2})\notin
T_{[W_2VXY_2]}^{3\delta} \}\nonumber\\ E_{10}&=&E_0^c\cap E_1^c\cap
E_2^c\cap E_4^c\cap E_6^c\cap \{\exists \vec{w'_1}\neq\vec{w^*_1}:
l(\vec{w'_1})=l(\vec{w^*_1}) \ \text{and} \ (\vec{w'_1},\vec{v^*},
\vec{Y_1})\in T_{[W_1VY_1]}^{4|\mathcal{X}|\delta}\}\nonumber\\
E_{11}&=&E_0^c\cap E_1^c\cap E_3^c\cap E_5^c\cap E_7^c\cap \{\exists
\vec{w'_2}\neq\vec{w^*_2}: l(\vec{w'_2})=l(\vec{w^*_2}) \ \text{and} \
(\vec{w'_2},\vec{v^*}, \vec{Y_2})\in
T_{[W_2VY_2]}^{4|\mathcal{X}|\delta}\}\nonumber
\end{eqnarray}

Apparently, for any $\epsilon'$, for $n>n_1(\epsilon',\delta)$, $P(E_0)\leq \epsilon'$. We have also
\begin{eqnarray}
P(E_1)&\leq&P(\vec{X}\in T_{[X]}^{\delta})P(\{\forall \ \vec{v}\in
\mathcal{C}_v,\ (\vec{X},\vec{v})\notin
T_{[XV]}^{2\delta}\}|\vec{X}\in T_{[X]}^{\delta})\nonumber\\
&\leq&\sum_{\vec{x}\in
T_{[X]}^{\delta}}P_{X}(\vec{x})(1-2^{-n(I(X;V)+\lambda)})^{nR_1}\nonumber\\
&\leq&\exp(-2^{-n(I(X;V)+\lambda-R_{V})}),
\end{eqnarray}
where Property 1) of the typical sequences and $(1-x)^y<e^{-xy}$ are used. Thus $P(E_1)\rightarrow 0$, provided that $R_{V}> I(X;V)+\lambda$.

$P(E_4)$ and $P(E_5)$ both tends to zero due to the Markov lemma; it
requires the condition $(\vec{v^*},\vec{X})\in T_{[VX]}^{2\delta}$ to
hold, which is indeed so given $E_1$ does not happen. Similarly, both
$P(E_8)$ and $P(E_9)$ tends to zero for the same reason. Notice that
if $(\vec{v^*},\vec{X},\vec{Y_1})\in T_{[VXY_1]}^{2\delta}$, then
$(\vec{v^*},\vec{Y_1})\in T_{[VY_1]}^{3|\mathcal{X}|\delta}$, thus
$\vec{v^*}$ can be correctly decoded if there is no other codewords in
the same bin satisfying the typicality test.

Conditioned on $E_1^c$, we have $(\vec{X},\vec{v})\in T_{[XV]}^{2\delta}$. Thus 
\begin{eqnarray}
P(E_2)&\leq& \sum_{(\vec{x},\vec{v})\in
T_{[XV]}^{2\delta}}Pr(\vec{x},\vec{v})(1-2^{-n(I(X;W_1|V)+\lambda)})^{nR_2}\nonumber\\
&\leq&\exp(-2^{-n(I(X;W_1|V)+\lambda_2-R_2)})
\end{eqnarray}
where property 2) of the typical sequences is used. Thus $P(E_2)$
tends to zero provided $R_{W_1}>I(X;W_1|V)+\lambda_1$. Similarly
$P(E_3')$ tends to zero provided $R_{W_2}>I(X;W_2|V)+\lambda_2$.

Conditioned on $E_1^c$, $\vec{y_1}\in T_{[Y_1]}^{\delta}$, since
codeword in $\mathcal{C}_v$ are generated independently according to
$P_U(\cdot)$
\begin{eqnarray}
P(E_6)&\leq& \sum_{\vec{v}\in \mathcal{C}_v}
2^{-nR_A}2^{-n(I(Y_1;V)-\lambda_1)}\nonumber\\
&=&2^{n(R_V-R_A-I(Y_1;V)+\lambda_1)}
\end{eqnarray}
where we have used property 2) of the typical sequences and the fact
the bin to which $\vec{v}$ is assigned is independent. Thus
$P(E_6)\rightarrow 0$ provided that
$R_A>R_V-I(Y_1;V)+\lambda_3$. Similarly $P(E_7)\rightarrow 0$ provided
that $R_A+R_A'>R_V-I(Y_2;V)+\lambda_4$.

Conditioned on $E_4^c$, $(\vec{v^*},\vec{Y_1})\in T_{[VY_1]}^{2|\mathcal{X}|\delta}$. Thus 
\begin{eqnarray}
P(E_{10})&\leq& 2^{nR_{W_1}}2^{-nR_B}2^{-n(I(Y_1;W_1|V)-\lambda_3)}\nonumber\\
&=&2^{n(R_{W_1}-R_B-I(Y_1;W_1|V)+\lambda_3)}
\end{eqnarray}
where property 3) of the typical sequences is used. Thus $P(E_{10})$
tends to zero provided
$R_B>R_{W_1}-I(Y_1;W_1|V)+\lambda_5$. Similarly, $P(E_{11})$ tends to
zero provided $R_C>R_{W_2}-I(Y_2;W_2|V)+\lambda_6$. Thus the rates
only need to satisfy
\begin{eqnarray}
&R_1=R_A+R_B>I(X;VW_1|Y_1)+\lambda'\\
&R_1+R_2=R_A+R_A'+R_B+R_C>I(X;VW_2|Y_2)+I(X;W_2|VY_1)+\lambda''
\end{eqnarray}
 where $\lambda'$ and $\lambda''$ are both small positive quantities
 and vanish as $\delta\rightarrow 0$ and $n\rightarrow \infty$; then
 $P_e\leq \sum_{i=0}^{11}P(E_i)\rightarrow 0$.  It only remains to
 show that the distortions constraints are satisfied as well. When no
 error occurs, then $(\hat{\vec{W_1}},\vec{X},\vec{Y_1})\in
 T_{[W_1XY]}^{3|\mathcal{V}|\delta}$ and
 $(\hat{\vec{W_2}},\vec{X},\vec{Y_1})\in
 T_{[W_2XY]}^{3|\mathcal{V}|\delta}$. By standard argument using the
 definition of the typical sequences, it can be shown that
\begin{eqnarray}
d(\vec{x},\hat{\vec{x_1}})\leq \Expt d[X,f_1(W_1,Y_1)]+\epsilon'
\end{eqnarray}
where
$\epsilon'=\max(d(x,\hat{x}))(3|\mathcal{V}\times\mathcal{W}_1\times\mathcal{X}\times\mathcal{Y}_1|\delta+P_e)$. Thus
the distortion can be made arbitrarily small by choosing sufficiently
small $\delta$ and sufficiently large $n$. Similar arguments holds for
the second stage decoder. This completes the proof.\hfill\QED



\section{Proof of the Theorem \ref{theorem:outer}}
\label{appendix:theorem3}

Assume the existence of $(n,M_1,M_2,D_1,D_2)$ RD SI-scalable code,
there exist encoding and decoding functions $\phi_i$ and $\psi_i$ for
$1=1,2$. Denote $\phi_i(X^n)$ as $T_i$. $\vec{X}_k^-$ will be used to
denote the vector $(X_1,X_2,\ldots ,X_{k-1})$ and $\vec{X}_k^+$ to denote
$(X_{k+1},X_{k+2},\ldots ,X_{n})$; the subscript $k$ will be dropped when
it is clear from the context. The proof follows the same line as the
converse proof in \cite{HeegardBerger:85}.  The following chain of
inequalities is standard (see page 440 of \cite{CoverThomas}). Here we omit the small positive quantity $\epsilon$ for simplicity.
\begin{eqnarray}
nR_1&\geq &H(T_1) \geq H(T_1|\vec{Y_1})=I(\vec{X};T_1|\vec{Y_1})=\sum_{k=1}^nI(X_k;T_1|\vec{Y_1}\vec{X}_k^-)\nonumber\\
&=&\sum_{k=1}^nH(X_k|\vec{Y_1}\vec{X}_k^-)-H(X_k|T_1\vec{Y_1}\vec{X}_k^-)\nonumber\\
&=&\sum_{k=1}^nH(X_k|Y_{1,k})-H(X_k|T_1\vec{Y_1}\vec{X}_k^-)\nonumber\\
&\geq&\sum_{k=1}^nI(X_k;T_1\vec{Y^-_1}\vec{Y^+_1}|Y_k).
\end{eqnarray}

Next we bound the sum rate as follows
\begin{eqnarray}
n(R_1+R_2)&\geq& H(T_1T_2)\geq H(T_1T_2|\vec{Y_2})=I(\vec{X};T_1T_2|\vec{Y_2})\nonumber\\
&=&I(\vec{X};T_1T_2\vec{Y_1}|\vec{Y_2})-I(\vec{X};\vec{Y_1}|T_1T_2\vec{Y_2})\nonumber\\
&=&\sum_{k=1}^n[I(X_k;T_1T_2\vec{Y_1}|\vec{Y_2}\vec{X}^-)-I(\vec{X};Y_{1,k}|T_1T_2\vec{Y_2}\vec{Y^-_1})].\nonumber
\end{eqnarray}
Since $(X_k,Y_{2,k})$ is independent of $(\vec{X}^-,\vec{Y^-_2},\vec{Y^+_2})$, we have
\begin{eqnarray}
I(X_k;T_1T_2\vec{Y_1}|\vec{Y_2}\vec{X}^-)=I(X_k;T_1T_2\vec{Y_1}\vec{Y^-_2}\vec{Y^+_2}\vec{X}^-|Y_{2,k})
\geq I(X_k;T_1T_2\vec{Y_1}\vec{Y^-_2}\vec{Y^+_2}|Y_{2,k})
\end{eqnarray}
The Markov condition $Y_{1,k}\leftrightarrow(X_k,Y_{2,k})\leftrightarrow(\vec{X}^-\vec{X}^+T_1T_2\vec{Y^-_1}\vec{Y^-_2}\vec{Y^+_2})$ gives 
\begin{eqnarray}
I(\vec{X};Y_{1,k}|T_1T_2\vec{Y_2}\vec{Y^-_1})=I(X_k;Y_{1,k}|T_1T_2\vec{Y_2}\vec{Y^-_1}).
\end{eqnarray}
Thus we have 
\begin{eqnarray}
n(R_1+R_2)&\geq&\sum_{k=1}^n[I(X_k;T_1T_2\vec{Y_1}\vec{Y^-_2}\vec{Y^+_2}|Y_{2,k})-I(X_k;Y_{1,k}|T_1T_2\vec{Y_2}\vec{Y^-_1})]\nonumber\\
&=&\sum_{k=1}^n[I(X_k;T_1T_2\vec{Y^-_1}\vec{Y_2}^-\vec{Y^+_2}|Y_{2,k})+I(X_k;\vec{Y^+_1}|T_1T_2\vec{Y_2}\vec{Y^-_1}Y_{1,k})].
\end{eqnarray}
The degradedness gives $Y_{2,k}\leftrightarrow Y_{1,k}\leftrightarrow (X_k,T_1T_2,\vec{Y^-_1}\vec{Y^-_2}\vec{Y^+_2})$, which implies
\begin{eqnarray}
n(R_1+R_2)&\geq&
\sum_{k=1}^n[I(X_k;T_1T_2\vec{Y^-_2}\vec{Y^+_2}\vec{Y^-_1}|Y_{2,k})+I(X_k;\vec{Y^+_1}|T_1T_2\vec{Y^-_2}\vec{Y^+_2}\vec{Y^-_1}Y_{1,k})].
\end{eqnarray}
Define $W_{1,k}=(T_1\vec{Y^-_1}\vec{Y^+_1})$ and $W_{2,k}=(T_1T_2\vec{Y^-_2}\vec{Y^+_2}\vec{Y^-_1})$, by which we have 
\begin{eqnarray}
nR_1&\geq&\sum_{k=1}^nI(X_k;W_{1,k}|Y_{1,k})\\
n(R_1+R_2)&\geq&\sum_{k=1}^n[I(X_k;W_{2,k}|Y_{2,k})+I(X_k;W_{1,k}|W_{2,k}Y_{1,k})].
\end{eqnarray}
Therefore the Markov condition $(W_{1,k},W_{2,k})\leftrightarrow X_k
\leftrightarrow Y_{1,k} \leftrightarrow Y_{2,k}$ is true. Next
introduce the time sharing random variable $Q$, which is independent
of the multisource, and uniformly distributed over $I_n$. Define
$W_j=(W_{j,Q},Q), \ j=1,2$. The existence of function $f_j$ follows by
defining
\begin{eqnarray}
f_1(W_1,Y_1)&=&\psi_{1,Q}(\phi_1(\vec{X}),\vec{Y_1})\\
f_2(W_2,Y_2)&=&\psi_{2,Q}(\phi_1(\vec{X}),\phi_2(\vec{X}),\vec{Y_2})
\end{eqnarray}
which leads the fulfillment of the distortion constraints.
It only remains to show both the bound can be written in single letter
form in $W_1,W_2$, which is straightforward following the approach in
(page 435 of) \cite{CoverThomas}. This completes the proof for
$\mathcal{R}_{out}(D_1,D_2)\supseteq\mathcal{R}(D_1,D_2)$. \hfill
$\blacksquare$

\section*{Acknowledgement}
The discussion with Emre Telatar is gratefully acknowledged. 

\bibliographystyle{IEEEtran}
\end{document}